\documentclass{aa}
\usepackage{graphicx}
\usepackage{natbib}
\bibpunct{(}{)}{;}{a}{}{,} 
\usepackage{array}
\usepackage{longtable}
\usepackage{textcomp}
\usepackage{multirow}
\usepackage{url}
\usepackage{multirow}
\usepackage[english]{babel}

\begin{document}
\title{The Connection between Supernova Remnants and the Galactic Magnetic Field: A Global Radio Study of the Axisymmetric Sample}

\author{J. L. West \inst{1} \and S. Safi-Harb
\thanks{Canada Research Chair}  
\inst{1} \and T. Jaffe \inst{2,3} \and R. Kothes \inst{4} \and T. L. Landecker \inst{4} \and T. Foster \inst{5}}

\institute{Dept of Physics and Astronomy, University of Manitoba, Winnipeg R3T 2N2, Canada\\
\email{jennifer.west@umanitoba.ca; samar.safi-harb@umanitoba.ca}
\and Universit\'e de Toulouse; UPS-OMP; IRAP;  Toulouse, France
\and CNRS; IRAP; 9 Av. colonel Roche, BP 44346, F-31028 Toulouse cedex 4, France
\and National Research Council, Herzberg Programs in Astronomy \& Astrophysics, Dominion Radio Astrophysical Observatory, PO Box 248, Penticton, V2A 6J9, Canada
\and Dept of Physics \& Astronomy, Brandon University, Brandon R7A 6A9, Canada}

\date{Received 23 July 2015 / Accepted 27 October 2015}

\titlerunning{Connection between SNRs and the Galactic Magnetic Field}
\authorrunning{West et al.}

\abstract{The study of supernova remnants (SNRs) is fundamental to understanding the chemical enrichment and magnetism in galaxies, including our own
Milky Way. In an effort to understand the connection between the morphology of SNRs and the Galactic magnetic field (GMF), we have examined the radio
images of all known SNRs in our Galaxy and compiled a large sample that have an "axisymmetric" morphology, which we define to mean SNRs with a "bilateral" or "barrel" -shaped morphology, in addition to one-sided shells. We selected the cleanest examples and model each of these at their appropriate Galactic position using two GMF models, those of Jansson \& Farrar (2012a), which includes a vertical halo component, and Sun et al. (2008) that is oriented entirely parallel to the plane. Since the magnitude and relative orientation of the magnetic field changes with distance from the sun, we analyse a range of distances, from 0.5 to 10~kpc in each case. Using a physically motivated model of a SNR expanding into the ambient GMF, we find the models using Jansson \& Farrar (2012a) are able to reproduce observed morphologies of many SNRs in our sample. These results strongly support the presence of an off-plane, vertical component to the GMF, and the importance of the Galactic field on SNR morphology. Our approach also provides a potential new method for determining distances to SNRs, or conversely, distances to features in the large-scale GMF if SNR distances are known.}

\keywords{ISM: supernova remnants -- ISM: magnetic fields -- radio continuum: ISM}
\maketitle

\section{Introduction}
Supernova explosions are some of the most significant and transformative
events in our Universe. Understanding supernova remnants (SNRs), the
remains of these explosions, is fundamental to understanding the chemical
enrichment and magnetism in galaxies, including our own Milky Way.
Shaped by energetic supernova explosions in the circumstellar medium (CSM)
or the interstellar medium (ISM), SNRs provide a powerful tool to study
the intrinsic properties of the explosion and the environment in which
they are expanding. Since the radio emission from SNRs arises primarily from
synchrotron radiation of relativistic particles in the presence of
compressed magnetic fields, radio observations are particularly useful
to probe their magnetic fields and connection to the Galaxy (see
\citealp{2012SSRv..166..231R} and references therein for a review). 

A subclass of SNRs, referred to as the \textquotedblleft bilateral\textquotedblright{}
or \textquotedblleft barrel\textquotedblright -shaped, which are characterized
by a symmetry axis (thus also referred to as `axisymmetric' by some authors)
dividing two opposing limbs of radio emission,
have a distinctive morphology and several authors have proposed a
connection to the Galactic magnetic field (GMF) and cosmic ray density.
An historical view, first proposed by \citet[see also \citealt{Whiteoak:1968fn}]{vanderLaan:1962wf}, is that
a SNR expanding in a relatively uniform magnetic field will sweep up and compress
the field where the expansion is perpendicular to the field lines.
Regions where the field increases produce higher intensity radio synchrotron
emission and thus are responsible for the appearance of the bright
limbs. \citet{Kesteven:1987vt} suggested that the majority of SNRs
would have this barrel shape and that distorted remnants were the
result of an inhomogeneous ISM. Further to this, \citet{2007A&A...470..927O}
showed that asymmetries in bilateral SNRs can be explained by gradients
of ambient density or magnetic field strength. Thus we consider \textquotedblleft unilateral\textquotedblright{}
SNRs, ones observed with a single well-defined limb (e.g. G024.7--00.6)  or ones having two limbs
with one much brighter than the other (e.g. G127.1+00.5), to have their origin from the
same processes as the bilateral type. Therefore, we include as axisymmetric those objects where we can draw a
line across the image, which separates two approximately symmetric structural
forms, even if the intensity of those forms differs. As such, we include SNRs in
which the structure on one side of the symmetry axis is quite faint.

\citet{1998ApJ...493..781G} pointed out that some early studies (\citealt{Shaver:1982ws,Manchester:1987um,1989ApJ...338..963L,Whiteoak:1996tn})
concluded that there was no clear relationship between the angle of
the axis of bilateral symmetry and the Galactic plane (hereafter called
the bilateral axis angle, $\psi$). Gaensler's analysis used
a tightly constrained sample of bilateral SNRs to conclude that there
was a significant tendency for the bilateral axes of these SNRs to
be aligned with the Galactic plane. However, even with a somewhat
restrictive selection criterion that constrained the sample to 17
bilateral SNRs, some are still significantly misaligned. One prime
example is G327.6+14.6 (SN1006), which is perhaps the cleanest and
most distinct example of a SNR with bilateral symmetry, and which
is rotated almost 90\textdegree{} to the direction of the Galactic
plane. 

These previous studies were selective on the sample of axisymmetric
SNRs, and in the modelling did not account for an off-plane, vertical
component of the Galactic field. The work reported here scrutinizes radio
images of all known SNRs in our Galaxy providing a more objective
and complete sample of axisymmetric SNRs, and makes use of the most comprehensive
GMF model to date. In Section \ref{sec:Magnetic-fields-of},
we summarize the current understanding of the magnetic field in SNRs (Section
\ref{sub:Magnetic-fields-of-1}) and the Galaxy (Section \ref{sub:Magnetic-field-of-MW}).
In Section \ref{sec:Previous-modelling-of} we discuss previous modelling
of bilateral SNRs including the distribution of cosmic-ray electrons
(CREs), which is another factor in determining the morphology of SNRs.
Our process for selecting the sample of SNRs used for this study is
discussed in Section \ref{sec:Selection-of-the}. In Section \ref{sec:Description-of-our}
we describe our modelling, including an outline of the modelling procedure (Section \ref{sub:Modelling-procedure}) and the details of our SNR model
(Section \ref{sub:Supernova-Model}).
We compare our models to the data in Section \ref{sec:Comparison-of-the-models-to-data}
and discuss features that appear in the models in Section \ref{sub:Features-of-the-models}.
In Section \ref{sec:Discussion}, we present our results including
a discussion about distance constraints (Section \ref{sub:Distance-constraints}), a comparison to an alternate magnetic field model (Section \ref{sub:Magnetic-fields-comparison}) and discussion about the magnetic fields of the SNRs (Section \ref{sub:Magnetic-fields-SNRs}).
Conclusions and suggestions for future work are presented in Section
\ref{sec:Conclusions-and-Future}.

\section{\label{sec:Magnetic-fields-of}Magnetic fields of SNRs and the Galaxy}

\subsection{\label{sub:Magnetic-fields-of-1}Magnetic fields of SNRs}

Observationally, one can find examples of shell-type SNRs with magnetic
fields oriented both radially (e.g. G120.1+01.4 (Tycho), G111.7--02.1
(CasA)) and tangentially (e.g. G119.5+10.2 (CTA-1) and G065.1+00.6)
(see \citealp{2012SSRv..166..231R}). \citeauthor{Milne:1987wi}\textquoteright s
atlas of SNR magnetic fields (1987) includes magnetic field maps of
15 shell-type SNRs. Of these seven have clear radial fields, four
have clear tangential fields, and the other four have fields that
cannot be classified as either radial or tangential. There are also
cases where both radial and tangential field geometries can be observed
in the same SNR. For example, observations by \citet{Reynoso:2013tr}
of G327.6+14.6 (SN1006) show that the field appears predominantly
radial at least within the interior of the SNR. These observations
also show evidence for a tangential field running along the outer
edge of the shell. \citet{Kothes:2001fj} show that G011.2--0.3 also
has a similar magnetic field structure with a radial interior and
tangetial appearance at the edge of the shell.

\citet{Milne:1987wi} first suggested that radial fields are a property
of young SNRs. The implication is that the magnetic field transitions
to a tangential geometry as the SNR ages when more ambient material
is swept up, and the magnetic field lines become more compressed.
One explanation given for the presence of radial fields in young SNRs
is that turbulence leads to selective amplification of the radial
component of the magnetic field (\citealt{Inoue:2013el}), although
\citet{2012SSRv..166..231R} state that the origin of radial fields,
particularly those observed immediately at the remnant edges, remains
unclear.

If the model of the compressed ambient field is correct, then the orientation of a SNR should be an excellent tracer
of the direction of the GMF. \citet{Whiteoak:1968fn} showed that
a magnetic field viewed from the side (i.e. completely perpendicular
to the line-of-sight) produced a tangential magnetic field with bilateral
appearance and when viewed \textquotedblleft end-on\textquotedblright{}
(i.e. completely parallel to the line-of-sight) produced a radial
magnetic field with circular appearance. We would expect to observe
SNRs in all orientations, and thus, if this model is true, there should
exist at least some cases where an observed radial field can be attributed
to the viewing angle rather than the youth of the SNR.

\subsection{\label{sub:Magnetic-field-of-MW}Magnetic field of the Milky Way
Galaxy }

In this work, we consider SNRs in the global context of the Galactic
magnetic field. Much work has been done in recent years on the global
magnetic field of the Galaxy (e.g. \citet{Brown:2002vc,Page:2007ce,Sun:2008bw,Sun:2009jh,Sun:2010gk,Pshirkov:2011gk,VanEck:2011hk,Jaffe:wl};
\citet[hereafter JF12]{Jansson:2012ep}; and \citet{Jansson:2012hl}. See also \citet{2015ASSL..407..483H} for a review.)
through studies of the rotation measures (RMs) of background sources
and modelling of the Galactic synchrotron radiation. Observations
using the RMs of extragalactic sources (\citealt{Brown:2001jt,VanEck:2011hk})
imply the presence of reversals, which are abrupt changes in the direction
of the large scale magnetic field of the Galaxy. 

Most current models and observations do not provide much information
about the \textit{vertical} component of the Milky Way's magnetic
field. However, observations of nearby, edge-on galaxies reveal an
\textquotedblleft X-shaped\textquotedblright{} halo component in all
cases studied thus far \citep{2009Ap&SS.320...77B,Beck:2013gp,Krause:2015if}. Analysis
of observations of the North Polar Spur also indicate the need for
a vertically oriented component to explain the RM signature of the
Spur \citep{2015ApJ...811...40S}. 

Of the above models, JF12 is the most recent Galactic field model that was systematically fit to data. This is the only model to date that includes a vertically oriented halo component. 
We also considered an alternate model by \citet{Sun:2008bw} for this study. This model uses
a magnetic field in the disk that has a constant pitch angle and uniform strength in azimuth, with reversals, and a toroidal halo component, but lacks any vertical component. As discussed in detail in Section \ref{sub:Magnetic-fields-comparison},
we find the JF12 model gives overall superior fits to our sample. We therefore choose
to use JF12 for our final analysis.

The JF12 model does have limitations and we note that aspects of it are not fully physically motivated. For example, the description of the large-scale regular field as a ring plus eight logarithmic spirals with discrete jumps in field strength between the spirals, and the distinction between a "disk component" and a "toroidal component", with the very different scale heights of the two components are both somewhat unphysical assumptions. Despite the limitations, we still consider this model is a reasonable choice for the present work. The discrete jumps in field strength in the spiral arms will not have a significant impact on our conclusions as the qualitative SNR morphology does not depend on the total field strength. Rather, it is the {\em orientation} of the field, which depends on the pitch angle and the relative strengths of the components, that determine the morphology of the SNR. The JF12 model describes the global, large-scale component of the Galactic field, which is our focus for this first study; i.e. whether
the large-scale, regular component dominates in the cases of clearly defined SNR shell limbs. While the morphology of individual SNRs may be affected by the specific features of the GMF model, taken as a whole, useful conclusions can still be drawn. 

The effect of the turbulent component on the global morphology of SNRs must also be considered as it is thought to be about twice as strong as the regular component. However, the turbulent power spectrum shows a scale of only a few parsecs within the spiral arms of the Galaxy (although this scale goes up to ~100 pc in the halo; \citealp{2015ASSL..407..483H}), and  
since nearly all SNRs are confined to the disk, most likely in the spiral arms\footnote{G327.6+14.6, 
has the highest latitude and is thought to be relatively nearby at a distance of 1.6-2.2 kpc  \citep{Ferrand:2012cr} and thus a height of about 400-550 pc above the plane,
or very close to the disk-halo boundary.}, and most are much larger than a few parsecs (our model SNRs are 40-60 pc in diameter, see Table \ref{tab:phys-ang-size}), we would expect the field to be dominated by the regular component in most cases. Not all SNRs are well-described by our model and this may be due to the turbulent component. The effect of turbulence on SNRs shapes, particularly if they are young and small, or if they are located in inter-arm regions where the turbulence scale can be large \citep[e.g.][]{1989ApJ...343..760R, 1993MNRAS.262..953O}, could be significant in some cases, however investigating the effect of this is beyond the scope of this work.

\subsubsection{\label{sub:Details-of-the}Details of the chosen Galactic magnetic
field model}

The large-scale regular field of JF12 is comprised of a disk component, a toroidal halo component,
and an out-of-plane, X-shaped, halo component. The field is set to zero for $r>20$~kpc and for a 1~kpc radius sphere centred on the Galactic centre.
See Figure~\ref{fig:JF12-B-field} for an illustration and Appendix \ref{sub:Coordinate-System} for a description of the coordinate system. See also Section 5.1 of JF12 for a detailed description of the model and its parameters.  

The disk component is purely in the $X$-$Y$ plane and includes a molecular
ring between 3~kpc and 5~kpc (Galacto-centric radius) that is purely
azimuthal with a constant field strength, $b_{ring}=0.1\,\mu G$.
Beyond 5~kpc, there are eight logarithmic spiral regions at radii:
5.1, 6.3, 7.1, 8.3, 9.8, 11.4, 12.7, and 15.5~kpc (the radii where
the spiral arm boundaries cross the negative $X$-axis).
The disk field extent is symmetrical with respect to the mid-plane and transitions to
the toroidal halo field at a height of $\sim$0.40~kpc.

The toroidal component is a purely azimuthal halo component that is characterized by separate field amplitudes in the north ($B_{n}=1.4\,\mu G$, with a transition radius 
$r_{n}=9.22$~kpc) and south ($B_{s}=-1.1\,\mu G$, with a transition radius $r_{s}>16.7$~kpc),
and with a vertical scale height of $\sim$~5.3~kpc (see Eq. 6 of JF12).

The out-of-plane halo component is described by an `X-shaped' field, primarily motivated by radio observations of haloes in external
edge-on galaxies \citep{2009Ap&SS.320...77B,Beck:2013gp,Krause:2015if}. This component is axisymmetric and has a poloidal shape, lacking any azimuthal form since this is included in
the toroidal field component.
The `X-field' component takes the form:
\begin{equation}
\begin{aligned}
& {B_{r,X-field}}={B_{X-field}}\cos\left({\Theta_{X-field}}\right)\left\{ \begin{array}{l}
1,{\rm {}}Z>0\\
-1,{\rm {}}Z<0
\end{array}\right.
\\
& {B_{Z,X-field}}={B_{X-field}}\sin\left({\Theta_{X-field}}\right)
\end{aligned}
\end{equation}

where $\Theta_{X-field}$ is the elevation angle, which is a function of
radius and $\Theta_{X-field}=90\text{\textdegree}$ at $r=0$ and $\Theta_{X-field}=49\text{\textdegree}$
at $r=4.8$~kpc, and $B_{X-field}=4.6\,\mu G$ (field strength of the X-field
component at the origin).

The total GMF is then the sum of the three components 
 (as illustrated in Figure~\ref{fig:JF12-B-field}):\\
\begin{equation}
\begin{aligned}
& {B_r}={B_{r,disk}}+B_{r,X-field} 
\\
& {B_{\phi}}={B_{\phi,disk}}+{B_{\phi,tor}}
\\
& {B_{Z}}={B_{Z,X-field}}
\end{aligned}
\end{equation}

\section{\label{sec:Previous-modelling-of}Previous modelling of Bilateral
SNRs }

It has been suggested that the observed morphology of bilateral SNRs
is greatly influenced by the distribution of the CREs (e.g. \citealt{Petruk:2009bg,2011A&A...531A.129B,Reynoso:2013tr}).
Two alternative distributions are often considered: the distributions for so-called
\textquotedblleft quasi-parallel\textquotedblright{} and \textquotedblleft quasi-perpendicular\textquotedblright{}
shocks (see \citealp{Jokipii:1982jy,1989ApJ...338..963L,Fulbright:1990gu}
and references therein), which show distinct observed morphologies
given the same magnetic field configuration. In particular, the axis
of bilateral symmetry of the radio synchrotron emission is rotated
by 90\textdegree{} between the two cases. In the quasi-perpendicular
case, the shock-normal is perpendicular to the ambient field, but
the axis of bilateral symmetry is aligned with the ambient field. In the quasi-parallel case the opposite is true; i.e. the shock-normal
is parallel to the ambient field, while the axis of bilateral symmetry
is perpendicular to the ambient field. 

For the case of SN1006 in particular, the matter of which distribution
is correct is still debated, with some arguing for the quasi-perpendicular
scenario (e.g. \citealt{Petruk:2009bg,Schneiter:2010dh}) and others
arguing for the quasi-parallel scenario 
(e.g. \citealt{Rothenflug:2004ch,2011A&A...531A.129B,Schneiter:2015fc}).

For the purposes of this study, we choose to use an isotropic distribution where the CREs are distributed uniformly in a shell. This will reveal the SNR morphology as if it were dependent solely on the compressed magnetic field. The overall shape of the morphology of the radio synchrotron emission in the quasi-perpendicular case, which despite it being more physically motivated than the isotropic case, is qualitatively the same than the isotropic case in a shell (see Figure~\ref{fig:CRE-geometry} and also \citealp{Fulbright:1990gu} and references therein). Although quantitatively these two cases are different, the goal of this study is a {\em qualitative} analysis to show whether the morphology of SNRs obtained from the compressed GMF alone is consistent with the observed morphology in a large sample. We intend to investigate the difference between the quasi-parallel and quasi-perpendicular cases in a future study.

We account for complexities
such as varying Galactic latitude, which will introduce a projected
component that is \textit{perpendicular} to the line-of-sight, and the effect of an intrinsic vertical GMF component. We find that these 
additions are very important for a global analysis of the morphology of axisymmetric SNRs.

We also investigate the global effects on SNRs in the context of the
GMF at varying longitudes. Using a global sample
of SNRs there is the suggestion of a correlation between the bilateral
axis angle, $\psi$, and the orientation of the Galactic
magnetic field. This relationship has the potential to reveal important
information on the Galactic field as \citet{2009IAUS..259...75K}
have previously suggested. For example, the presence of a reversal
will imply an abrupt change in $\psi$.
This is because the field model includes sign flips in the spiral component
of the disk field, which alone would not change the morphology of
the SNR, but when added to the smooth toroidal component, the total field 
orientation (not just the sign) changes. This in turn
implies an abrupt change in $\psi$. This also gives a potential new method 
to place constraints on distances
to SNRs; in particular those at Galactic longitudes where the direction
of the magnetic field, and thus the model SNR morphology, is changing
rapidly along the line-of sight. SNRs could also be used to place
constraints on distances to features, such as reversals, in the GMF.

\section{\label{sec:Selection-of-the}Selection of the SNRs}

For this paper we are studying the extent to which the SNR morphology
and radio synchrotron emission are related to the regular component
of the GMF that has been compressed by the SNR shock
wave. Hence, we wish to include only the cleanest and most clearly-defined
shells where the shell morphology is more likely not to be significantly
altered through interaction with local enhancements in surrounding
gas, turbulence in the GMF, nor influenced by
a pulsar or a pulsar-wind nebula (PWN). In order to examine the appearance
of the SNRs, we have searched the literature and data archives to
collect the best-available radio images of all SNRs, excluding the
pure shell-less PWNe; i.e. those that are classified as any type other
than the filled-centre type as defined by \citet{Ferrand:2012cr}%
\footnote{SNRcat: \url{http://www.physics.umanitoba.ca/snr/SNRcat/}%
} and \citet{Green:2014vb}. This is a total of 293 objects. We compile all radio images in a companion website, Supernova remnant Models \& Images at Radio Frequencies (SMIRF)%
\footnote{SMIRF: \url{http://www.physics.umanitoba.ca/snr/smirf/}}. In Table~\ref{tab:Number-of-SNRs}
we summarize the numbers of SNRs of each type. This table represents
the up-to-date numbers at the time of this writing; however the website
is dynamic and classifications and exact counts of SNRs of various
types are subject to change. 

Thermal composite-type (also called mixed-morphology), plerionic composite-type, and ``unknown''-type
SNRs are considered, but with caution since these objects may involve
more complex physics due to the presence of a central compact object
and/or local enhancements of the ISM. 

The thermal composite-type SNRs exhibit thermal X-ray emission in
the SNR interior, but lack the shell seen at radio wavelengths. While
the mechanisms responsible for their X-ray emission are still not
fully understood, these remnants are generally found to be interacting
with nearby dense interstellar structures, thus complicating our modelling
and analysis. There are five thermal composites that have a very clearly
defined axisymmetric morphology: G021.8--00.6, G116.9+00.2, G166.0+04.3,
and G359.1--00.5, plus G093.3+06.9 that is also labeled as a plerionic
composite (see Table~\ref{tab:Number-of-SNRs} for a summary of the
numbers of thermal composite SNRs of each type).

The plerionic composite-type may be influenced by the central compact
object and surrounding nebula. These are included only in cases
where we can be convinced that the shell is large compared to the PWN and the central
pulsar is far enough away from the shell walls that we judge its
influence to be negligible. Only one plerionic composite-type SNR is
considered to be very clearly defined: G119.5+10.2 (plus G093.3+06.9
mentioned above). 

We also looked at SNRs with unknown type. None of these are clearly defined, but four have possible, though
poorly defined axisymmetric morphology. A notable one in this category
is G039.7--02.0 (W50), which could possibly be an example of an axisymmetric
type with jets from the binary source SS433.

There are 199 SNRs that are defined as shell-type \citep{Ferrand:2012cr}.
Based on the images that we reviewed, the morphology of 112 are not consistent with what we call axisymmetric (i.e. a clear double- or single-sided shell), and we do not consider them further\footnote{These are labeled as "Not axisymmetric" in Table \ref{tab:Number-of-SNRs}}. Some of these non-axisymmetric type SNRs have a completely
undefined structure, while others have a defined, yet filamentary
structure, which have significant emission through the centre of the
shell instead of just at the edges. Still others have a ring-like
or round appearance (there are 15 out of 293 that we consider to have a very clearly defined or somewhat defined round appearance). These may prove to be interesting SNRs to analyze
in future work (see Section \ref{sec:Magnetic-fields-of}), but
we disregard them for the present analysis. This leaves 87 out of
the 199 shell-type SNRs that we classify as axisymmetric. 

In total, we have a sample of 112 SNRs with an axisymmetric (or possibly
axisymmetric) morphology, including the selected thermal composite,
plerionic composite, unknown, and shell-type objects (see Table \ref{tab:Number-of-SNRs}). We assigned
a level of uncertainty to the classification using the labels: very
clearly defined (33 SNRs), somewhat defined (21 SNRs), and not clearly
defined (58 SNRs). All 112 SNRs have been modelled and those models
are available on the companion website (SMIRF: \url{www.physics.umanitoba.ca/snr/smirf/}). The analysis in this
paper will only include the 33 SNRs with very clearly defined, ``clean''
morphology. 

\section{\label{sec:Description-of-our}Description of our model}

\subsection{\label{sub:Modelling-procedure}Outline of the modelling procedure}

We use the Hammurabi code (Appendix \ref{sub:Hammurabi-Code}) to model the Stokes I, Stokes Q, and Stokes U synchrotron emission (defined below) at the coordinates of each of our selected SNRs for eleven discrete distances: 0.5, 1, 2, 3, 4, 5, 6, 7, 8, 9, \& 10
kpc. The SNR is modelled as a spherical shock or "bubble", which deforms the field lines of the JF12 GMF model at the SNR location (see Section \ref{sub:Supernova-Model}). Since the Galactic field model varies with distance, the model
SNRs show differences in morphology as a function of distance as well.
This potentially provides a constraint on the distance of an SNR based
on its morphology. For each SNR position and distance, the modelling
process is as follows: 
\begin{enumerate}
\item Run Hammurabi to output the JF12 model of the GMF
and the NE2001 thermal electron density for the direction of a SNR
on a particular line-of-sight (i.e. for a particular set of Galactic
coordinates). The field is written with a resolution of 1 pc per voxel.
The total size of the volume is determined by the assumed radius of
the model SNR and the distance. Table~\ref{tab:phys-ang-size} summarizes
the physical and angular sizes for the SNRs bubbles modelled at the
various distances. These sizes were
chosen to be consistent with the approximate average size expected
for an SNR in the Sedov phase. At nearer distances, a smaller size
is chosen to reduce computation time (reduces the overall size of
the volume) and reduce the angular size to something that is consistent
with the observations (the SNR with the largest angular scale in our sample is G315.1+02.7,
which has a size of 190$^{\prime}$). 

Note that the integration that computes the total
Stokes I, Stokes Q, and Stokes U parameters is not done at this stage.
Rather, Hammurabi is used at this stage only to write the magnetic field components and thermal electron density
at each point along a particular line-of-sight to a file.
\item Using Matlab, read the portion of the magnetic field and thermal electron
density data files (output from Hammurabi) at the position of the
SNR and apply the numerically defined coordinate transformation function
(see Section \ref{sub:Supernova-Model}) to these voxels. These sections
of the files are then overwritten with the newly calculated transformed
magnetic field and thermal electron density values. 
\item Run Hammurabi a second time. This time the magnetic field and thermal electron
density data files are passed as input to Hammurabi. In addition,
an analytical model of the distribution of CRE
is defined within the code. The ambient CRE density is defined as
in Appendix \ref{sub:Hammurabi-Code} while the CRE within the shell
of the SNR is compressed uniformly around the shell as the thermal
electrons. Hammurabi then numerically integrates and produces output
for the model Stokes I, Q, and U parameters, which is also described
in Appendix \ref{sub:Hammurabi-Code}. 
\end{enumerate}

\subsection{\label{sub:Supernova-Model}Supernova Remnant Model}
After the initial supernova explosion, the conventional description
is that the SNR is ejecta-dominated until the swept-up
mass exceeds the mass of the ejecta. It is at this point, roughly
1000 years after the explosion (depending on the progenitor,
explosion energy, and ambient density), that the blast wave can be described using a
Sedov-Taylor solution (\citealp{1959sdmm.book.....S,Korobeinikov:1991vf}).
This solution has the advantage of being self-similar
with a well-defined analytical form that describes the thermal electron mass density as
a function of radius (see Figure~\ref{fig:coord-transform}). We model
the thermal electron density and magnetic field of a spherical shell compressed
by a Sedov-Taylor blast wave. 
We note that some SNRs in our sample are either quite young (e.g. G001.9+00.3)  or possibly in a radiative phase of expansion (e.g. the thermal composites), and as a result, are likely not in the Sedov-Taylor phase. We argue however that the ambient medium is still compressed enough that their morphology can be approximated by using the Sedov-Taylor density model. Furthermore, we are here mostly concerned with their overall {\it morphology} at radio wavelengths, rather than an absolute measurement of their emission in radio or other wavelengths. 

In order to compute the magnetic field in the shell, we assume
that the magnetic field vectors are
\textquotedblleft frozen-in\textquotedblright{} to the ambient plasma,
which is a common assumption for ionized plasma.
We then adopt the method of \citet{Franzmann:2014tr}, who developed a coordinate transformation technique to model
the magnetic field in molecular cloud cores. 

The transformation, outlined in Appendix \ref{sub:Coordinate-Transformation}, takes an initial 3D magnetic field, $\bf{B}$, which is compressed by a spherical shock that "drags" the field lines and gives the appropriately transformed magnetic field, $\bf{B'}$. While the initial thermal electron mass density is assumed to be constant in the region surrounding the SNR, the initial magnetic field is not required to be uniform (as illustrated in the example shown in Figure \ref{fig:coord-transform}); it can have an arbitrary distribution. The SNR can thus be modelled in the context of the Galaxy by using this method to transform the model GMF and "insert" an SNR at a particular location.

The process of using this coordinate transformation to alter the magnetic field results in an output that is compressed at the edges, and we measure the magnitude of the compressed field to be around six times greater that the initial field (see Figure \ref{fig:coord-transform}). An amplification factor of six agrees with that derived by \citet{vanderLaan:1962wf} for the case of spherical geometry (see Eq. 57). The magnetic field is not compressed at locations where the shock normal is oriented parallel to the magnetic field as illustrated in Figure \ref{fig:coord-transform}. Recent X-ray observations show that the magnetic field is amplified by a much larger factor than this, which is most likely due to local turbulent acceleration that amplifies the already compressed field (\citealp{Uchiyama:2007dx,Uchiyama:2008ef,2012SSRv..166..231R}). We do not take this additional amplification into account, but note that this would impact the {\em quantitative} result, and not the {\em qualitative} morphology that is the focus of this work. 

\section{\label{sec:Comparison-of-the-models-to-data}Comparison of the models
to data}

We present results of modelling 33 SNRs distributed around the Galaxy
(See Figure~\ref{fig:Galaxy-map} and Appendix~\ref{sub:Models-Data}). In order
to compare the data to the model we use two parameters. The first
parameter is the angle $\psi$, which we remind the reader is the
projected angle between the axis of bilateral symmetry and the Galactic plane.
The second parameter is the ratio of the peak brightness level between
the two sides. For the models,
the angle, $\psi$ was found by measuring the mean value of the pixels
along a line extending across the diameter of the model SNR. The line
was rotated in 1\textdegree{} increments and the position where the
mean of the pixels along the line has a minimum value was taken as the angle through the symmetry
axis. The real data images are not as clean as the model images as
they contain background emission, point sources, and noise. Thus we
did not obtain good results measuring the angle $\psi$ on the real
data using this method and it was found that a by-eye method resulted
in a better measurement. Therefore we measured $\psi$ by using
a rectangular box sized to fit the gap between the lobes. We rotated the box
in 1\textdegree{} increments and the best angle
was determined based on how well it apparently divides the emission
into two distinct lobes. We estimate the uncertainty of this measurement
to be \textpm 5\textdegree . This estimate is based on the variation
resulting from repeated independent measurements.

After the angle is determined, a circular region with 8 pie-slice
shape wedges is defined, centred on the SNR (this procedure was done
for both the models and the data). Two of the slices are centred on
the symmetry axis, which means two other slices should be centred
on the bilateral \textquotedblleft lobes\textquotedblright . We then
found the ratio of the mean brightness in the two wedges for the SNR
lobes on opposite sides (i.e. mean north limb divided by mean south
limb). This ratio was measured for all of the models using the automatically
determined best-fit angle and for the data using the by-eye best-fit
angle in each case. 

If the SNR were perfectly symmetrically bilateral the ratio should
be 1. If this ratio is~>~1 it means the SNR/model is brighter in the
north, and if this ratio is~<~1 it means it is brighter in the south.
We determined the best-fit model overall by looking first at the best
fitting angles and then comparing the ratios. If two models had an
angle with an equally good fit, then the ratio was used to select
the best model overall. These results are summarized in Table~\ref{tab:data-vs-model}.

\subsection{\label{sub:Features-of-the-models}Features of the models}

As shown in Appendix~\ref{sub:Models-Data}, most of the models have a symmetric,
"bilateral'' appearance with two well separated limbs of uniform
brightness. This is because the magnetic field is relatively uniform at the particular location where the SNR was modelled and
thus the compressed field is more or less equal on both sides. However
some of the models have a circular morphology or other unusual features.

In some cases, the models have a "round"
or circular appearance, for example, the case of G036.6+02.6 at $d=5$~kpc (see Appendix~\ref{sub:Models-Data}). This is due to the magnetic field
being primarily directed along the line-of-sight at those locations
(i.e. the vectors are coming directly at the observer or pointing
directly away from the observer). It should be noted that these "round"
models should be intrinsically fainter since the observed
synchrotron emission depends on the {\em perpendicular} component
of the magnetic field. This brightness difference is not
apparent in the figures since the model images have been normalized
for the purpose of comparing them to the observed SNR morphology.

In other locations, we find that the model SNRs have one limb significantly
brighter than the other; for example, G028.6--00.1 at distances of
2--4~kpc (see Appendix~\ref{sub:Models-Data}). This brightness
difference can be explained by asymmetries in the magnetic field model.
In some cases, if the field is changing, there can be a difference
between the two limbs where the field has a larger line-of-sight component
on one side and a larger perpendicular component on the other. In
this case the limb with the stronger perpendicular component will
be brighter. In other cases, the magnitude of the field is stronger
on one side compared to the other, which will also result in an asymmetry
in the brightness.

For some models, some sharp and dark features appear in the images.
For example, for the case of G317.3--00.2, the model images up to 6
kpc show such a feature. These apparent lines are the result of sharp
transitions in the magnetic field model and can be emphasized by the
vector addition of the various components of the model (see Section
\ref{sub:Details-of-the}). These transitions will appear sharper
and somewhat unphysical in these models. It is possible that such
transitions are present in the real Galactic field although they would
likely be smoother and less abrupt. 

In Figure~\ref{fig:Top:-Models-at} we show several models at the
position of G317.3--00.2 at a distance of 2~kpc to show the impact
of excluding the various field components. In this example, 
the X shape is contributing a line-of-sight component, in addition to a vertical GMF component. Thus, the model that includes
the X-field shows strong asymmetry with the southern limb of the
SNR being much brighter, which is consistent with the data. Figure~\ref{fig:Top:-Models-at} shows that for this particular example, the asymmetry is
due primarily to the inclusion of the X-field and illustrates
the impact of including this component.

\section{\label{sec:Discussion}Results and Discussion}

We find that 25 out of the 33 SNRs have an angle that agrees with
the angle from the Galactic model within 10\textdegree{} (see Figure
\ref{fig:Histogram-of-the}). When we compare the brightness, we find
that for 23 out of 33 SNRs the models and data agree in the sense
that they are brighter/fainter on the same side (i.e. both would be
bright in the North or both bright in the South). In five of the cases
the brightness difference is border-line where both the model and
data have values close to one (i.e. equally bright on both sides). There
are five cases (G021.8--0.6, G046.8--00.3, G054.4--00.3, G166.0+04.3, G327.4+1.0)
where there is a significant disagreement between the brightness distribution
between the model and the data in terms of this measurement. It is important to note that at least two of these cases (G021.8--0.6 and G166.0+04.3) are known to be thermal composites
interacting with a molecular cloud.

The fact that nearly 75\% of our `clean' sample of axisymmetric SNRs is well modelled with the JF12 model provides strong
support for the impact of the GMF on SNR morphology, as well as the need for a vertical component for Galactic field models (see also Section 7.2).
We remind the reader that we had purposely selected the clean sample in our investigation in order to minimize
observational bias based on poor-quality data.

We did a preliminary review of the other 113 SNRs in the axisymmetric sample (see Section \ref{sec:Selection-of-the})
to compare the models and data based on a visual inspection. Even though in many cases it is difficult to judge the fit due to poor quality data, we find that there are intriguing matches for a number of SNRs between model and data that are good prospective case studies.
The models for all 113 SNRs can be reviewed on the companion website (SMIRF: \url{www.physics.umanitoba.ca/snr/smirf/}).

Nearly 25\% of our sample is clearly not well fitted by the Galactic field model, implying that other effects are in play.
Aside from local electron acceleration effects (e.g associated with quasi-parallel/perpendicular mechanisms) and turbulence that are not accounted for in this work,
additional factors in affecting the SNR morphology include the SN progenitor and expansion into the CSM.
This is particularly expected for the youngest SNRs that are still under the influence of the progenitor's mass-loss history (e.g. \citet{1982ApJ...259L..85C}).
Significant departures from the standard Sedov-Taylor evolutionary phase can also arise from the SNR expansion into stellar wind bubbles blown by the pre-supernova progenitor.
In fact, it has been argued that SNRs resulting from explosions of very massive stars can spend a significant fraction of their lifetime in their progenitor bubbles, 
which would then affect their evolution and morphology for tens of thousands of years (see e.g. \citealt{2005ApJ...630..892D,2011MmSAI..82..781D}). 
Therefore, they interact with the ISM at a much later stage in comparison to the type Ia remnants expanding normally in a less complex CSM.
A systematic study of the SNR sample taking into account known age and classification (Ia vs core-collapse)
will help us address this question. It is also possible that in some cases the progenitor bubble itself is carved by the GMF.
Interesting case studies include G296.5+10.0 whose striking bilateral morphology has been suggested to be affected by a magnetized progenitor wind
(Harvey-Smith et al. 2010). Overall, we can successfully model the morphology of an impressively large number of axisymmetric SNRs  and while these other effects
can have a significant bearing on the appearance of a SNR, the GMF still seems to be the dominant factor.

\subsection{\label{sub:Distance-constraints}Distance constraints}

Depending on the longitude of the SNR, the orientation (pitch angle) of the GMF model, and how the model varies
along the line of sight for that direction, the models can fairly
tightly constrain the distance in some cases (for example G003.7--00.2, G093.3+06.9, G296.5+10.0, G327.6+14.6, and G350.0--02.0); while in other cases,
a much larger range of distances could account for the observed emission. 
We remind the reader that there are significant uncertainties in the specific 
features of the JF12 model (see Section \ref{sub:Magnetic-field-of-MW}). Thus one must be cautious with interpreting these results for individual SNRs
 and dedicated case studies are necessary for constraining the distance uncertainties in specific instances. This is particularly true for SNRs near the Galactic centre where the GMF is more uncertain and may be 
 dominated by turbulent magnetic fields.

Our best fit distances are determined by comparing the angles and
brightness ratios of the data with the model. These are summarized
in Table~\ref{tab:data-vs-model}. Local variations in
the magnetic field could change the distance interpretation but we
expect that the large scale field would dominate in most cases, particularly
in these cases where the shell is clear and well-defined. 

When we compared the distances for the best fit of the model with
the distances given in the literature (and compiled in SNRcat, \citealp{Ferrand:2012cr})
we find that out of the 33 SNRs, 18 have some distance estimate
published in the literature and our results agree with 15 of the 18
distance estimates. These results are summarized in Table~\ref{tab:data-vs-model} (see also Figure~\ref{fig:Galaxy-map}). Only 3 SNRs have quite poor agreement: G065.1+0.6, G332.0+0.2 and
G359.1--00.5.

\textbf{G065.1+0.6:} The distance to the shell is estimated to be
9--9.6~kpc \citep{Tian:2006bp} whereas our best-fit model gives $5_{-2}^{+2}$
kpc. We do note that the distance to a nearby pulsar (PSR J1957+2831), possibly associated
with the shell, is estimated to be 7~kpc, which is in agreement with
the upper limit of our range. A further investigation of the distance to this remnant
and its association with the pulsar is needed.

\textbf{G332.0+0.2:} This SNR is estimated to be at > 6.6~kpc \citep{Caswell:1975tn},
which is a kinematic distance based on measurements of OH absorption.
This disagrees with our best-fit model distance of $1{}_{-1}^{+2}$
kpc. Given that the size of the radio shell is 12$^{\prime}$, at
a distance of 1~kpc, the shell's physical size would be 3.4 pc, which
would imply that the object is quite young. Further investigation and a new distance estimate
are required.

\textbf{G359.1--00.5:} The Suzaku X-ray study of this object \citep{Ohnishi:2011ev}
implies a distance close to the Galactic centre and the authors assume
a distance of 8.5~kpc, which is in agreement with other work that
puts the distance at 8--10.5~kpc \citep{Frail:2011vp,Uchida:1992je}.
Our best fit model puts the distance at $1_{-1}^{+2}$~kpc, but we
note that the JF12 model does not attempt to model the GMF right at the Galactic centre (see also Figure 5 showing that there are no model
fits for distances between 7 and 10~kpc, which bracket the distance from observations).\\

The fact that the distance agrees in the majority of cases supports the use of this model as a distance indicator in cases where the distance to the SNR is unknown. Conversely, quality imaging of SNRs, combined with
good distance information, can give valuable input to determination of distances to features of the GMF.

\subsection{\label{sub:Magnetic-fields-comparison}Comparison with \citet{Sun:2008bw}}

In addition to JF12, we computed models for all axisymmetric SNRs using the GMF model of \citet{Sun:2008bw}, which does $not$ include a $Z$-component. 
These two GMF models were both derived from fits to observations and the models both use the same spiral pitch angle. Since the field strength does not come into this analysis, the main difference between the two models is the geometry, namely the distances and directions of the reversals and the addition of the X-field in JF12. 
For each model we selected the distance that matches best, but in many cases, no good match was available, especially for the models using \citet{Sun:2008bw}. The
Figure~\ref{fig:Histogram-of-the} illustrates that the JF12 model provides a substantially better fit to the data, particularly
in terms of $\psi$, in nearly all cases. 

In our sample of 33 SNRs, there
are about nine cases where the Sun et al. (2008) model gives a reasonably
good fit. In most cases this occurs where the orientation of the symmetry
axis of the SNR is parallel to the Galactic plane and where the Galactic
latitude is small ($<|2{^\circ}|$). Figure~\ref{fig:Comparison-of-SNR}
shows a comparison of the two field models for three illustrative
SNRs: one at relatively high latitude, G296.5+10.0; one at a mid-range
latitude, G016.2--02.7; and one in the galactic plane, G046.8--00.3.

\subsection{\label{sub:Magnetic-fields-SNRs}Magnetic fields of the SNRs}

The model observations give us simulated Stokes Q and U polarization parameters. These are used to produce polarized intensity ($PI = \sqrt {{Q^2} + {U^2}}$) and polarization angle ($PA=\frac{1}{2}\tan^{ - 1}\frac{U}{Q}$) values that are used to make plots of the simulated magnetic field as in Figure~\ref{fig:Comparison-of-SNR} (where the PA gives the orientation of the electric field).

Of the 33 SNRs in our very well defined sample, 13 have a magnetic
field that has been observed through polarization studies. Of these,
8 have been observed to have a tangential magnetic field: G016.2--02.7,
G065.1+00.6, G093.3+06.9, G119.5+10.2, G127.1+00.5, G166.0+04.3, G182.4+04.3,
and G296.5+10.0. In every one of these cases, simulated magnetic field plots for the models also show a tangential magnetic field.
Two such cases, G016.2--2.7 and G296.5+10.0, are shown in Figure~\ref{fig:Comparison-of-SNR}. 

\citet{Roger:1988im} suggest that a vertically
oriented field may be responsible for the appearance of SNRs G296.5+10.0
and G327.6+14.6 (SN1006) and our study supports this conclusion. We
note that RM observations by \citet{2010ApJ...712.1157H} lead those
authors to conclude that G296.5+10.0 has a radial field, possibly
due to the progenitor star. However the observations have very poor
UV coverage and are only sensitive to smaller structures making this conclusion uncertain. Additionally, RM observations are sensitive only to the line-of-sight component of the magnetic field. 
We propose that a twisted vertical
field could explain both the RM result and still be consistent with
a vertically oriented tangential field as shown by \citet{Milne:1987wi}.

Three SNRs in our sample have what we term ``mixed'' magnetic fields,
where the field is not obviously tangential or radial. These are G021.8--00.6,
G054.4--00.3 and G116.9+00.2. Both G021.8--00.6 and G116.9+00.2
are thermal composite-type SNRs. In addition, two of these SNRs G021.8--00.6
and G054.4--00.3 were noted above as having poor fits in terms of the
brightness ratio, which is perhaps not surprising since they
seem to be more complex cases. G054.4--00.3 in particular is interesting
in that the morphology of the model bears a striking similarity to
the data despite the fact that the brightness ratios do not agree.
The simulated polarization vector plots for these SNRs (Figure~\ref{fig:Simulated-polarization-mixed-B})
also show deviations from a purely tangential field and thus would
also be considered to have ``mixed'' magnetic fields.

Only two SNRs in our sample have been observed to have radial fields:
G046.8--00.3 and G327.6+14.6 (SN1006). It is interesting that in the case of G046.8--00.3, the simulated polarization vector plot for the JF12 is
tangential, but the plot using the Sun et al. (2008) model at
the corresponding distance (4~kpc) does indeed show a radial magnetic
field (See Figure~\ref{fig:Comparison-of-SNR}).

G327.6+14.6 (SN1006) has also been observed to have a radial field
\citep{Reynoso:2013tr}, but the observations also reveal the suggestion
of a tangential field at the edges. In our model, the simulated polarization
vector plot for the JF12 field is tangential, but we note that
the field configuration at this location is close to being directed
primarily along the line of sight, which would produce a
radial field. This will be addressed in a future dedicated study.

\section{\label{sec:Conclusions-and-Future}Conclusions and Future Work}

We have focused in this paper on the cleanest and most complete sample
of Galactic axisymmetric SNRs, i.e. those showing an axis of symmetry
often referred to in the literature as \textquotedblleft bilateral\textquotedblright{}
or \textquotedblleft barrel\textquotedblright{}-shaped SNRs. We have modelled
these remnants using the comprehensive GMF model of Jansson
and Farrar (2012a, JF12), which takes into account for the first time an
off-plane, X-shaped component of the GMF that is motivated by observations of the magnetic field in external galaxies. 
We have made all of these images and models available on our companion website 
(SMIRF: \url{www.physics.umanitoba.ca/snr/smirf/}). We
stress it is not the details of the JF12 model that matter the most
for our modelling, but rather the presence of the vertical, X-shaped
component (described in detail in Section \ref{sub:Magnetic-field-of-MW})
that is not included in previous models. We have demonstrated that:

1. We can reproduce the observed morphologies of individual SNRs, in
particular the bilateral axis angle, through a simple and physically
motivated model of an SNR expanding into the ambient GMF. In addition to the morphology, the 
magnetic fields predicted by the models (i.e. tangential or radial) are consistent with the observed magnetic fields in nearly all cases.

2. If the large scale Galactic field is known, this method can
predict distances to SNRs, or conversely, if SNR distances are
known, they could constrain the large-scale GMF
at their position.

3.  A systematic comparison of a sample of SNRs at different positions
in the Galaxy in combination with this modeling method can distinguish
between two large-scale GMF models, even given
the uncertainties in the distances.

4. A large-scale GMF model without a vertical component
is not consistent with our sample of SNRs. Therefore, either the large-scale
Galactic field must have a vertical field component, or the
simple and physically motivated model of SNR expansion into the ambient
field is missing an element that must not only alter the morphologies
of individual SNRs (e.g. ambient density changes) but must do so in
a systematic way throughout the Galaxy. It is difficult to envision
such a mechanism that is as simple and natural (and motivated by observations
of external galaxies) as an X-shaped field component. 
Although it is suggestive, these results do not conclusively 
prove that the vertical field must have an X-shape. Nevertheless, this paper 
strongly supports the presence of a vertical halo component in the Milky Way Galaxy.

In future work, we plan to present a more detailed analysis of some
select SNRs with sufficient data in the radio (and in X-rays for the
non-thermal shells), comparing our results not only to the SNR morphology,
but also to the observed radio polarization and rotation measure observations
where possible. Furthermore, SNRs with a ring-like or round appearance
are also an interesting area for future study with this modelling
to investigate whether some of them may be a result of line-of-sight
magnetic fields. 

The model used in this study (JF12) included only the regular component
of the GMF, and did not include any turbulent component 
that would be associated with the large-scale Galactic field or local-to-the-SNR acceleration mechanisms.
In a future study we would like to investigate the impact of turbulence on SNR morphology. 
Since the sample of Galactic SNRs represents a range of different ages and thus different
scales, SNRs can be used as a tool to investigate the scale of the
turbulence. In particular, an SNR that is much smaller than the scale of
the GMF turbulence may see the turbulent component 
as a regular component. In that case the SNR may still show axisymmetric
morphology, but at an angle consistent with the turbulent magnetic
field (Gwenael Giacinti 2015, private communication). 
Thus a careful analysis of SNRs at various scales may give some important clues to
the outer scale of the turbulence spectrum in the GMF.

Other possible extensions of this work include investigating global
properties of the GMF such as the pitch angle
of the spiral pattern and the precise shape of the vertical field. Another exciting future
extension is to conduct an analysis of SNRs with well-constrained
distances from other robust methods, to estimate distances to features
in the GMF such as reversals and transitions.

\begin{acknowledgements}
This research was primarily supported by the Natural Sciences and
Engineering Research Council of Canada (NSERC) through a Canada Graduate
Scholarship (J. West) and the Canada Research Chairs and Discovery
Grants Programs (S. Safi-Harb). The modelling was performed on a local
computing cluster funded by the Canada Foundation for Innovation (CFI) and
the Manitoba Research and Innovation Fund (MRIF). We acknowledge and thank Gilles Ferrand for useful discussions
and contributions to SNRcat used in this study, and Erica Franzmann for her assistance with the coordinate transformation code.\\
\\

The radio images presented in this paper and companion website made
use of data obtained with the following facilities: CHIPASS 1.4 GHz
radio continuum map, Canadian Galactic Plane Survey (CGPS) and other
data from the Dominion Radio Astrophysical Observatory, Southern Galactic
Plane Survey, Effelsberg 100m Telescope and Stockert Galactic plane
survey (2720 MHz) (via MPIfR's Survey Sampler), Molonglo Observatory
Synthesis Telescope (MOST) Supernova Remnant Catalogue (843 MHz),
Sino-German 6 cm survey. Additionally we acknowledge the use of NASA's
SkyView facility (http://skyview.gsfc.nasa.gov) located at NASA Goddard
Space Flight Center for the data from the 4850 MHz Survey/GB6 Survey,
NRAO VLA Sky Survey, Sydney University Molonglo Sky Survey (843 MHz),
and the Westerbork Northern Sky Survey (325 MHz Continuum). Very Large
Array (VLA) data was acquired via the NRAO Science Data Archive and
the Multi-Array Galactic Plane Imaging Survey. We thank Michael Bietenholz
and David Kaplan for providing the VLA images of G021.6--00.8 and G042.8+00.6,
respectively. \\
\\
This research made use of APLpy, an open-source plotting package for Python (http://aplpy.github.com) and Astropy, a community-developed core Python package for Astronomy (Astropy Collaboration, 2013). This research also made use of Montage. It is funded by the National Science Foundation under Grant Number ACI-1440620, and was previously funded by the National Aeronautics and Space Administration's Earth Science Technology Office, Computation Technologies Project, under Cooperative Agreement Number NCC5-626 between NASA and the California Institute of Technology.\\
\\
This research has made use of the NASA Astrophysics Data System (ADS).\\
\\
We are grateful to and thank the referee for the helpful and thorough comments that significantly improved the paper.

\end{acknowledgements}

\bibliographystyle{aa}
\bibliography{references}

\begin{appendix}

\section{\label{sub:Coordinate-System}Coordinate System }

The Hammurabi code uses the conventional coordinate system of a top-down plot as viewed from above the North Galactic Pole.
In this view, the $X$-$Y$ plane represents the plane of the Galaxy,
and the $Z$-axis is directed perpendicular to the plane. Many Galactic
field models, including the models of \citet{Sun:2008bw} are pure
toroidal models that do not include an intrinsic $Z$-component (however
the magnitude of the $X$- and $Y$-components do depend on $Z$).
That is, all of the magnetic field vectors are parallel to the disk
of the Galaxy. 

When we observe in some arbitrary direction with Galactic
longitude, $l$, and latitude, $b$, it is more useful to consider
the component of the magnetic field that is along the line-of-sight,
which we call $x$, and the components in the plane of the sky,
which we will call $y$ and $z$ (see Figure~\ref{fig:JF12-B-field}).
We transform the coordinates from the $X$, $Y$, $Z$ cartesian coordinate
system to the $x$, $y$, and $z$ line-of-sight
coordinate system via the following rotation equations:
\begin{equation}
\begin{aligned}
& x=X\cos(b)\cos(l)+Y\cos(b)\sin(l)+Z\sin(b)
\\
& y=-X\sin(l)+Y\cos(l)
\\
& z=-X\sin(b)\cos(l)-Y\sin(b)\sin(l)+Z\cos(b)
\end{aligned}
\end{equation}

For observations at $b=0{^\circ}$, we are looking directly into the
Galactic plane and thus the line-of-sight coordinate, $x$,
is entirely in the plane and the $x$-$y$ plane is coincident
with the $X$-$Y$ plane. In this case, the $z$-component is equivalent
to the $Z$-component, which is intrinsic to the particular magnetic
field model being used (i.e. for \citet{Sun:2008bw}, $B_{Z}=B_{z}=0$,
but for JF12, $B_{Z}=B_{z}\ne0$). For non-zero Galactic latitudes,
the $x$-$y$ plane is tilted with respect to the $X$-$Y$
plane by the angle $b$. This introduces a projected $B_{z}-$component
that depends solely on the line-of-sight component. In particular,
for longitudes where the magnetic field is primarily along the line-of-sight
(e.g. $l=50{^\circ}$, see Figure~\ref{fig:JF12-B-field}) $B_{x}$
is maximum and $B_{y}$ is close to zero. The $B_{z}$-component
is exactly zero at $b=0{^\circ}$ but it increases rapidly as $|b|$
increases and the azimuthal component gets projected onto the plane
of the sky. For longitudes where the line-of-sight component is close
to zero (e.g. $l=170{^\circ}$ and $l=355{^\circ}$, see Figure~\ref{fig:JF12-B-field}),
$B_{x}$ and $B_{z}$ are close to zero and $B_{y}$
is a maximum. The $B_{y}$-component is perpendicular to the
rotation, and thus does not get projected.

\section{\label{sub:Hammurabi-Code}Hammurabi Code}

We use the Hammurabi code\footnote{http://sourceforge.net/projects/hammurabicode/} \citep{Waelkens:2009bn}, which has been used previously to model
the large-scale structure of the GMF. The Hammurabi
code models the synchrotron emission and Faraday Rotation given an
input 3-D magnetic field, thermal electron distribution, and CRE distribution. The code was modified to transform the components of a magnetic field to the line-of-sight coordinate frame, which is critical for this work (see Section \ref{sub:Coordinate-System} and \ref{sub:Modelling-procedure}).
We use the GMF model of JF12 and the thermal electron
density distribution defined by the NE2001 code (\citealp{Cordes:2002tt}),
which is the same model used in JF12. 

The CRE model defines the spectral index and the CRE spatial density
distribution at all points in the volume. These quantities are defined
separately for the region inside the model SNR bubble and for the
surrounding Galactic medium. For the ambient Galaxy, we use the power
law spectral index, $p=-3$ (defined as {\bf $dN/dE$ $\sim$ $E^p$}) as this is the typical value used in other
Galactic models and is the value adopted by JF12. For the spatial
density, we use the distribution from WMAP (\citealp{Page:2007ce})
since it is the default distribution available in Hammurabi. The ambient CRE distribution serves only to provide the
surrounding background emission. Since all available models vary smoothly on the scale of the
SNRs the choice of the specific model
will not impact the SNR morphology nor will it affect our conclusions.

For the SNR, the CRE distribution is scaled according to the thermal electron distribution.
That is, we assume the CRE density is compressed in the shell and
that this compression is spherically symmetric around the whole shell. As discussed in Section~\ref{sec:Previous-modelling-of}, we use
this assumption since we are investigating the role of the compressed ambient magnetic field on SNR morphology.
The qualitative morphology of the isotropic case is very similar
to the quasi-perpendicular CRE distribution since, in both cases,
the resulting model SNRs are bright around the equatorial belt and
faint at the polar caps (recall Figure~\ref{fig:CRE-geometry}). A quantitative analysis, and 
comparison to the quasi-parallel scenario is beyond the scope of this work, but will be investigated in future work.

The Hammurabi code uses the HEALPIX pixelization scheme \citep{Gorski:2005ku},
which divides the sky into pixels of equal areas. The angular resolution,
$\Delta\theta\approx\sqrt{\frac{3}{\pi}}\frac{{3600'}}{{N_{{\rm {SIDE}}}}}$,
is determined by setting the parameter $N_{{\rm {SIDE}}}$
(where $N_{{\rm {SIDE}}}$ is a power of 2).
We use $N_{{\rm {SIDE}}}$ = 8192, which corresponds to an angular resolution of
$~0.5'$. The step size along the line-of-sight, $\Delta r$
is set to 1 pc and the maximum distance along the line-of-sight, $r_{{\rm {max}}}$
is set to 1~kpc further than the distance to the SNR for a particular
model (for example for modelling a SNR at 4~kpc, we would set $r_{{\rm {max}}}=5$
kpc). This assumes that the SNR dominates the emission in any given
field and thus, the Galactic emission missing from behind the SNR
does not affect the analysis. 

Hammurabi calculates a number of quantities. For this work we analyze
the total radio synchrotron emission, Stokes I, and the polarization
vectors Stokes Q and Stokes U. These are expressed as \citep{Waelkens:2009bn}:
\begin{equation}
\begin{aligned}
& {I_{i}}={C_{I}}B_{i,\bot}^{\left({1-p}\right)/2}{\nu^{\left({1+p}\right)/2}}\Delta r
\\
& {P_{i}}={C_{P}}B_{i,\bot}^{\left({1-p}\right)/2}{\nu^{\left({1+p}\right)/2}}\Delta r
\\
& \Delta{RM_{i}}=0.81{n_{e}}{B_{i,\parallel}}\Delta r
\\
& {\chi_{i}}={\chi_{i,0}}+\sum\limits _{j=1}^{j=i}{{RM_{j}}{\lambda^{2}}}
\\
& {Q_{i}}={P_{i}}\cos\left({2{\chi_{i}}}\right)
\\
& {U_{i}}={P_{i}}\sin\left({2{\chi_{i}}}\right).
\end{aligned}
\end{equation}

Here, $i$ corresponds to the $i$-th volume element, $p$ is the
power law spectral index (see above), $C_{I}$ and $C_{P}$ are factors that are
dependent on $p$ (see \citet{Waelkens:2009bn,1979rpa..book.....R}),
$\nu$ is the frequency of observation, which for this work is set
to 1.4 GHz, $P_i$ is the polarized specific intensity, $RM$ is the rotation measure, $n_{e}$ is the thermal
electron density and $\lambda$ is the wavelength of observation (0.21 m corresponds to  1.4
GHz). The total Stokes $I$, Stokes $Q$ and Stokes
$U$ parameters are then found by summing the volume elements, $i$,
along the line-of-sight.

\section{\label{sub:Coordinate-Transformation} Coordinate Transformation}
A coordinate transformation is used to add the SNR into the GMF for a particular location. The assumption is that
a region of uniform thermal electron mass-density is transformed into a region with
a mass density described by some well defined profile, which in our
case is the Sedov-Taylor solution. Then, we can define two coordinate
systems that describe how the thermal electron mass is distributed in these two frames
and solve for the transformation that would convert from one frame
to the other. The initial thermal electron mass-density distribution
is uniform and the explosion occurs at a point $r=0$. In this frame,
the $r$-coordinate has uniform spacing on a numerical grid. Using
conservation of mass, a new coordinate system, called $r'$, is defined
by numerically integrating concentric spheres and comparing the mass
to the uniform system. The mass of the two systems is related by
\begin{equation}
\frac{4}{3}\pi\rho{r^{3}}=4\pi\int{\rho'{{r'}^{2}}dr'};
\end{equation}

where $\rho$ is the thermal electron mass density in the uniform system and $\rho'$
is the mass density in the transformed system (i.e. the spherical
shell compressed by a Sedov-Taylor blast wave).

The original uniformly distributed mass is rearranged to follow the
density function defined by a standard self-similar Sedov-Taylor solution.
Using the original $r$-coordinate (uniform density) and the new $r'$-coordinate
(Sedov density profile) we numerically solve for a coordinate transformation
matrix (Jacobian) that transforms $r$ to the new $r'$-coordinate
system that is given by
\begin{equation}
{\bf {J}}=\left[{\begin{array}{ccc}
{\frac{{\partial x'}}{{\partial x}}} & {\frac{{\partial x'}}{{\partial y}}} & {\frac{{\partial x'}}{{\partial z}}}\\
{\frac{{\partial y'}}{{\partial x}}} & {\frac{{\partial y'}}{{\partial y}}} & {\frac{{\partial y'}}{{\partial z}}}\\
{\frac{{\partial z'}}{{\partial x}}} & {\frac{{\partial z'}}{{\partial y}}} & {\frac{{\partial z'}}{{\partial z}}}
\end{array}}\right]
\end{equation}

(e.g. \citealp{boas2005mathematical}). This coordinate transformation
matrix can then be applied to transform the vector field (magnetic
field vectors) where
\begin{equation}
{\bf {B'}} = \frac{{\bf {J}}}{{\det{\bf {J}}}}{\bf {B}}
\end{equation}

\section{\label{sub:Models-Data}Data shown in comparison to the models. }
In each case the data are shown on the left (image references are summarized
in Table~\ref{tab:data-vs-model}). The angle of the green-coloured
box was determined visually and its angle represents the bilateral
axis of the data. The angle is measured from the horizontal and is
set to a positive value if it is in the first quadrant, and to a negative value if
it is in the second quadrant. To the right of the image is the strip
of models that were made for the position of the particular SNR and
at the various distances as labeled (in~kpc). In some cases the Galactic
field model is undefined at a location and so the model image will
show blank. This affects G001.9+00.3, G003.7--0.2, G354.8--00.8 and G359.1--00.5, which are blank due to the magnetic field model being zero in central region of the Galaxy and
G156.2+05.7, G166.0+04.3, and G182.4+04.3 due to the fact that
the thermal electron/CRE density model is set to zero for distances beyond
17~kpc from the Galactic centre. The set of best fitting models, based on the visual
appearance of the angle is highlighted with an orange box, while
the overall best fit model that takes all parameters into account
is indicated by a green line that is drawn at the angle $\psi$ that
was measured for that particular model. Where a published value for
the distance is available, the range is indicated by an arrow
above the models (references for these distances are summarized in
Table~\ref{tab:data-vs-model}).

\begin{figure*}
\centering
\includegraphics[width=17cm]{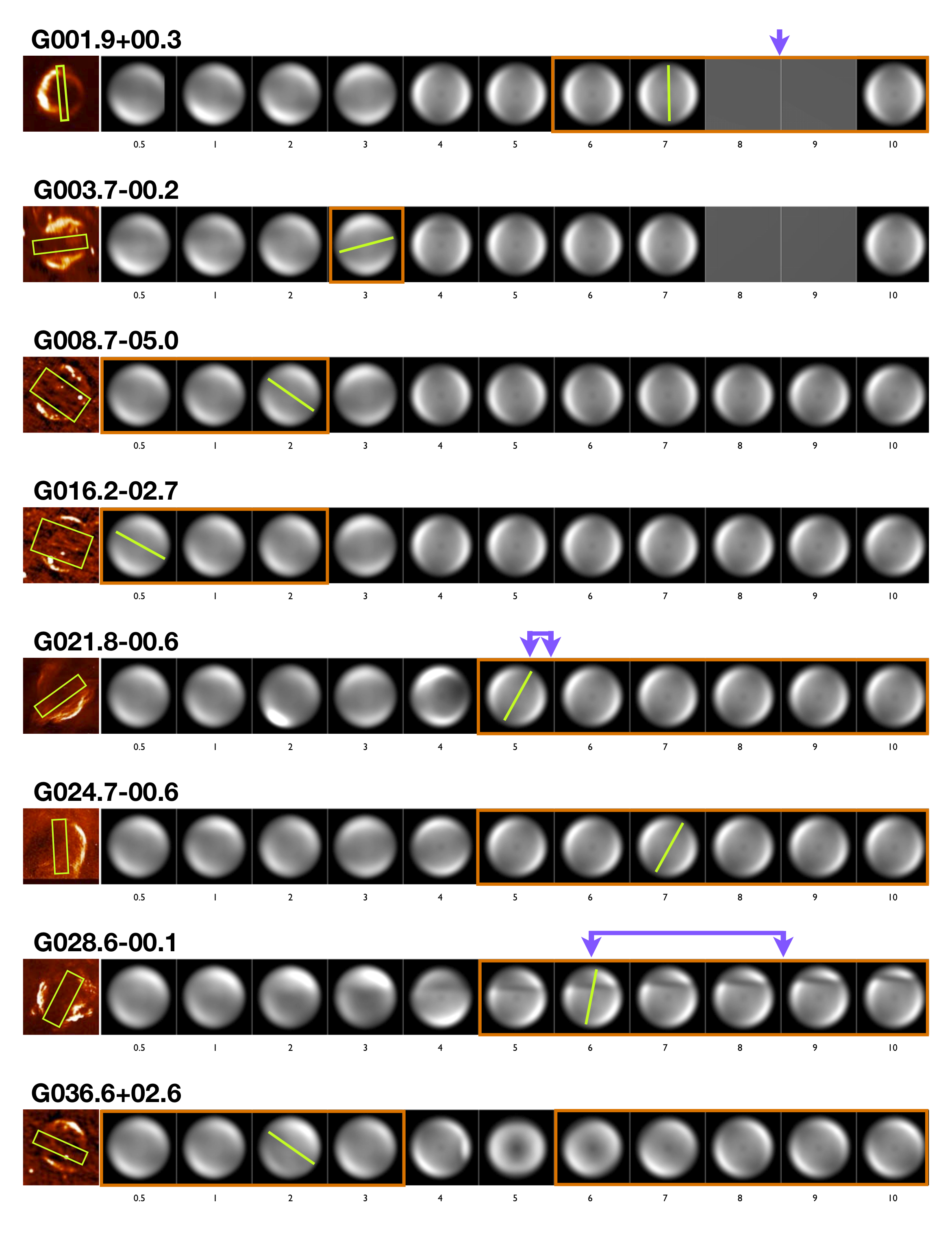}
\end{figure*}
\begin{figure*}
\centering
\includegraphics[width=17cm]{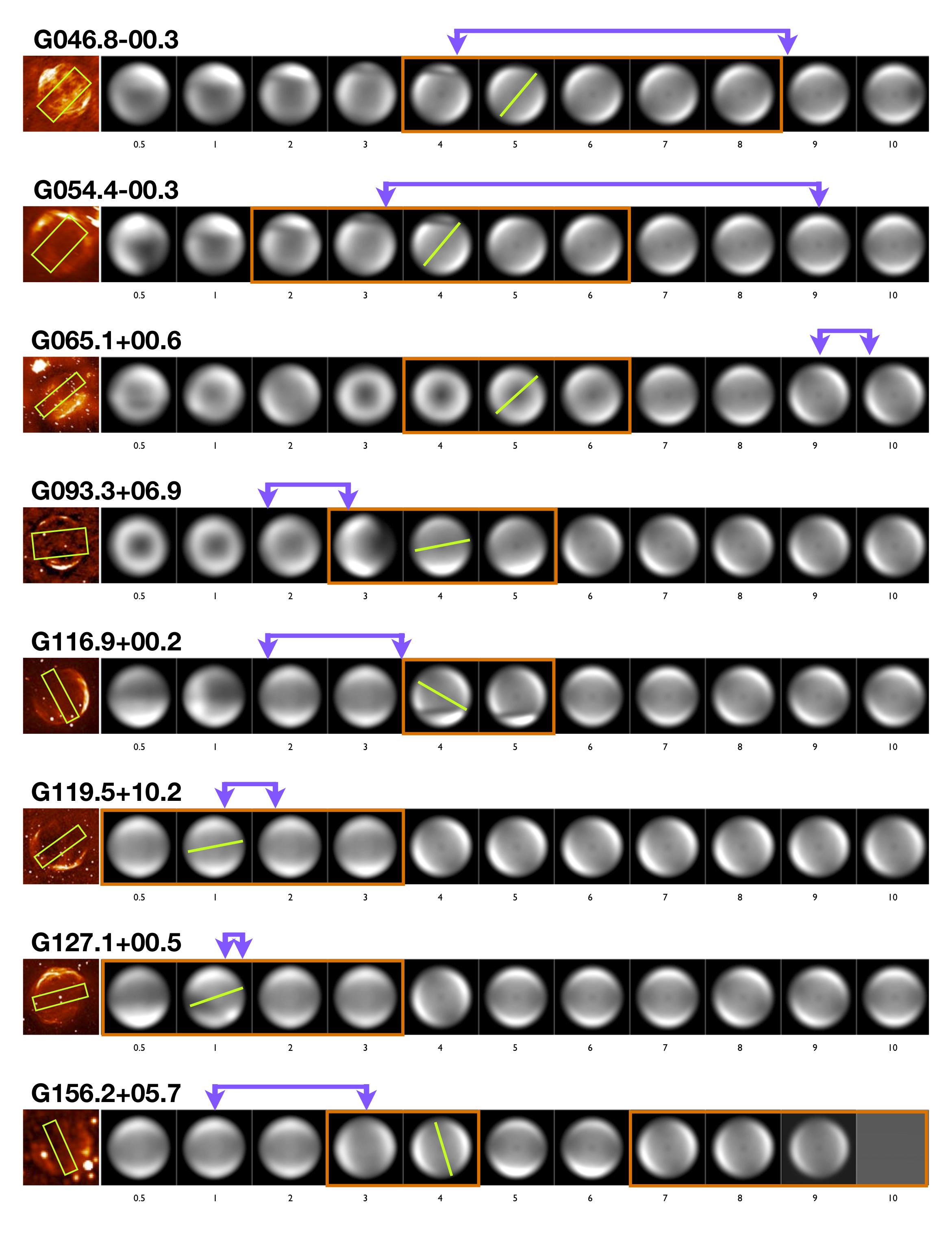}
\end{figure*}
\begin{figure*}
\centering
\includegraphics[width=17cm]{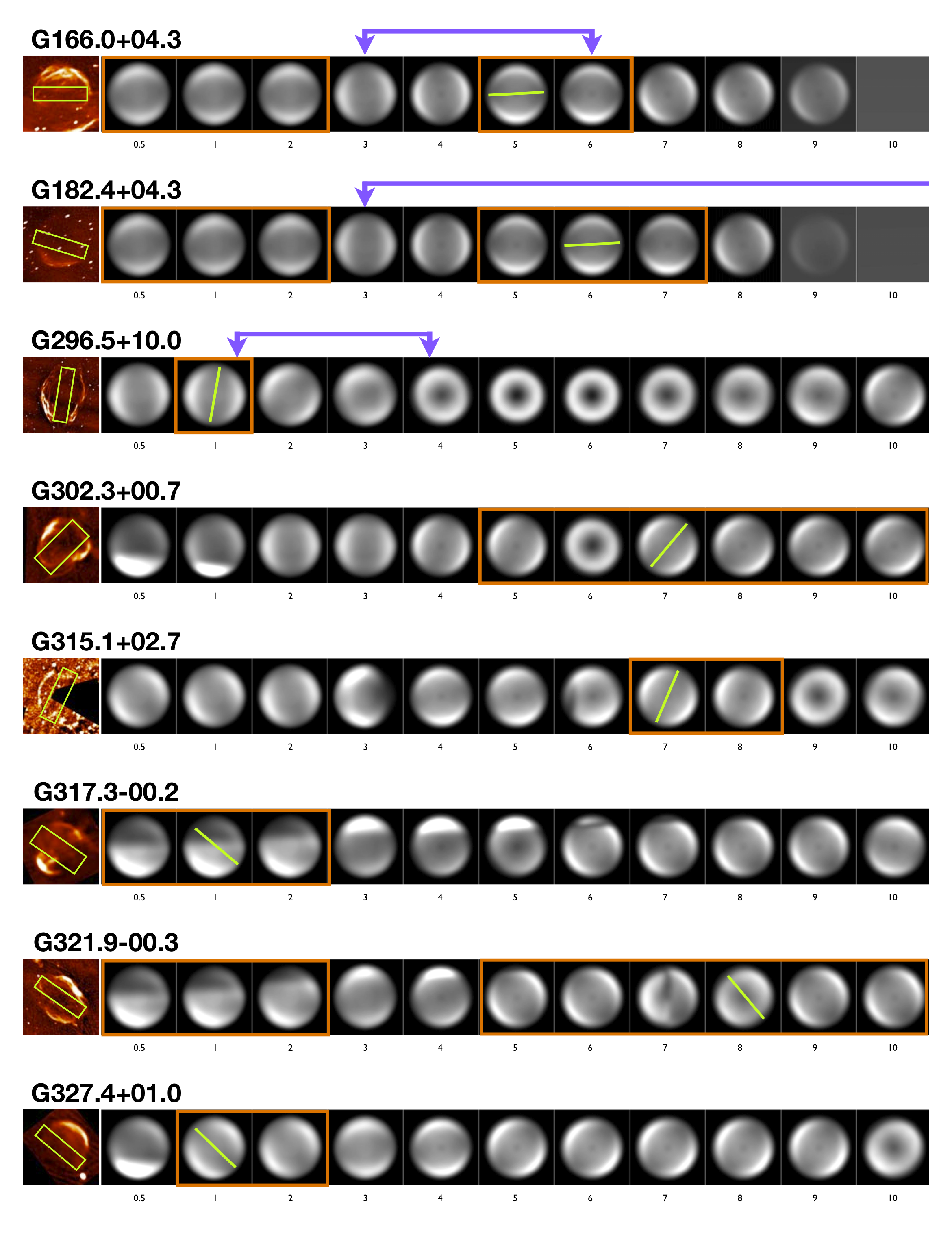}
\end{figure*}
\begin{figure*}
\centering
\includegraphics[width=17cm]{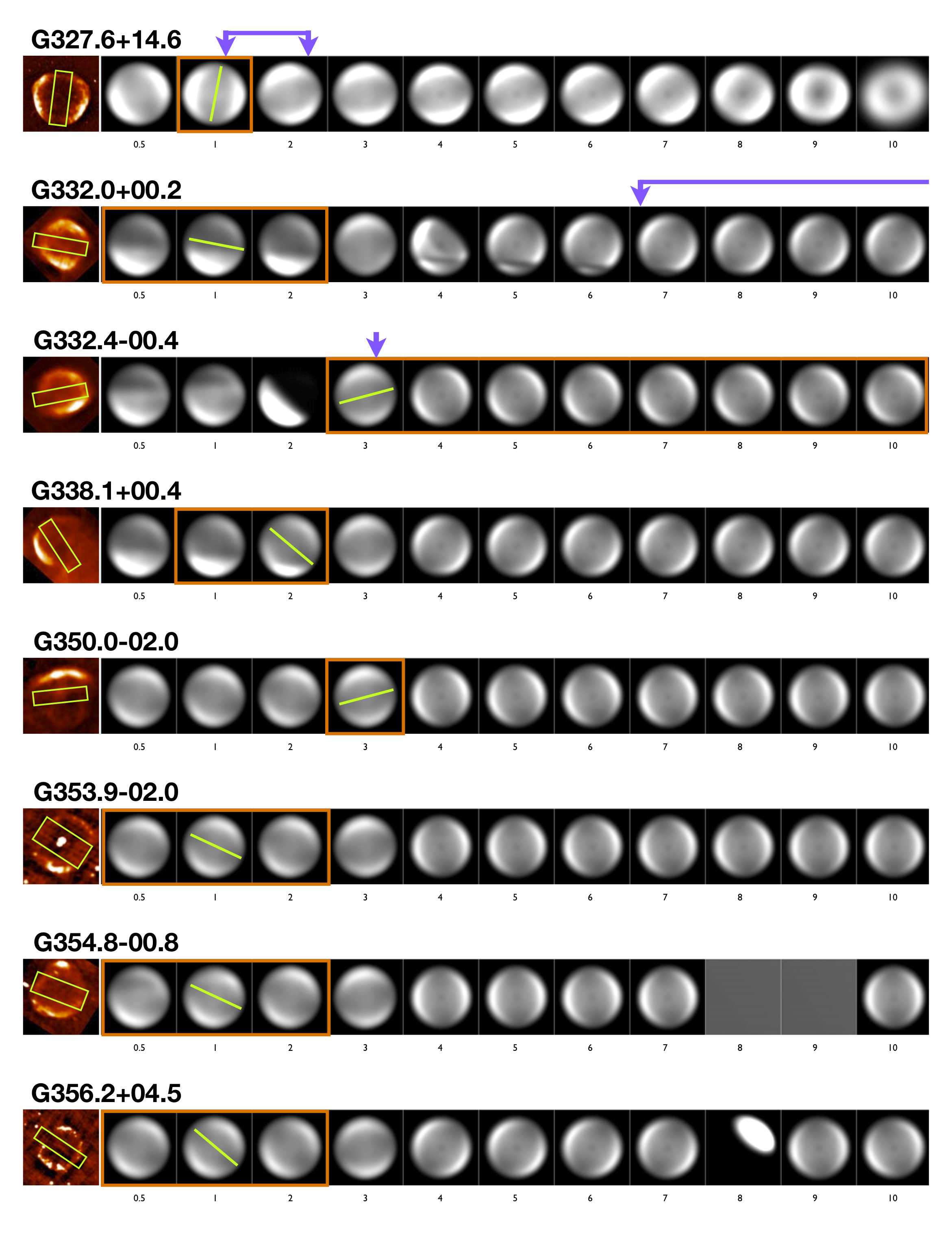}
\end{figure*}

\begin{figure*}
\centering
\includegraphics[width=17cm]{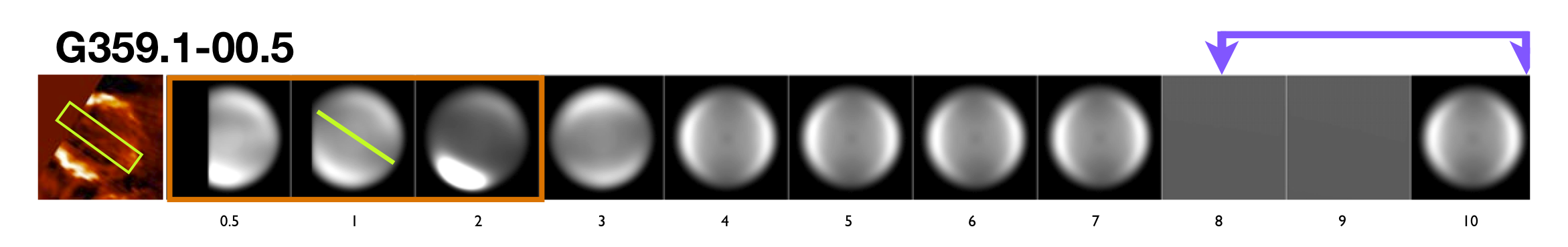}
\end{figure*}

\end{appendix}

\begin{figure*}
\centering
\includegraphics[width=12cm]{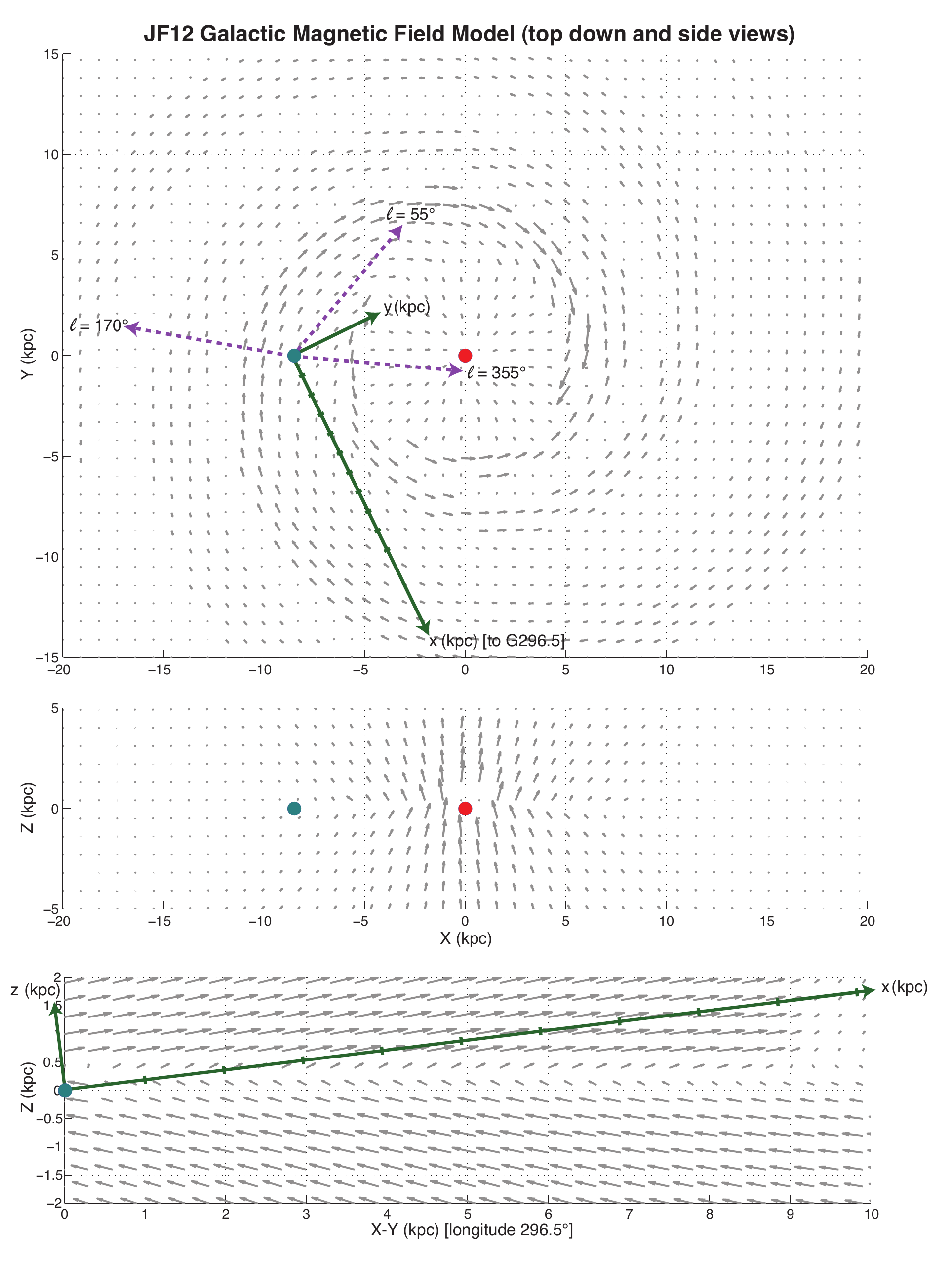}
\protect\caption{\label{fig:JF12-B-field}Plots of the magnetic field from JF12. {\bf Top
panel}: Top view of the $X$-$Y$ plane of the Galaxy cut through $Z=0$.
The green arrows show the $x$-$y$ coordinate system for SNR G296.5+10.0.
The dashes on the $x$-axis are marked at 1~kpc intervals. The purple,
dashed arrows show other longitudes, $l=55{^\circ}$, where the GMF
is primarily along the line-of-sight and $l=170{^\circ}$ and $l=355{^\circ}$,
where the GMF is primarily perpendicular to the line-of-sight.
The filled, red circle marks the Galactic centre and the filled, green
circle marks the position of the Sun. {\bf Centre panel}: $X$-$Z$ plane cut
through $Y=0$. As in the top panel, the filled, red circle marks
the Galactic centre and the filled, green circle marks the position
of the Sun. The shape of the ``X-field'' can be seen. {\bf Bottom panel}:
GMF lines are shown as cut along $l=296.5{^\circ}$. Here the
horizontal axis of this plot is in the $X$-$Y$ plane of the Galaxy along
$l=296.5{^\circ}$. The vertical axis of this plot shows the z-axis
of the Galaxy. As in the top panel, the green arrows show the primed
coordinate system, in this case the $x$-$z$ coordinates, for the case
of SNR G296.5+10.0, which has b=10.0\textdegree . The dashes on the
$x$-axis are marked at 1~kpc intervals as in the top panel. Here, one
can see that for $d=1$~kpc, the GMF vectors are nearly pointed
along the $z$-axis.}
\end{figure*}

\begin{figure*}
\centering
\includegraphics[width=12cm]{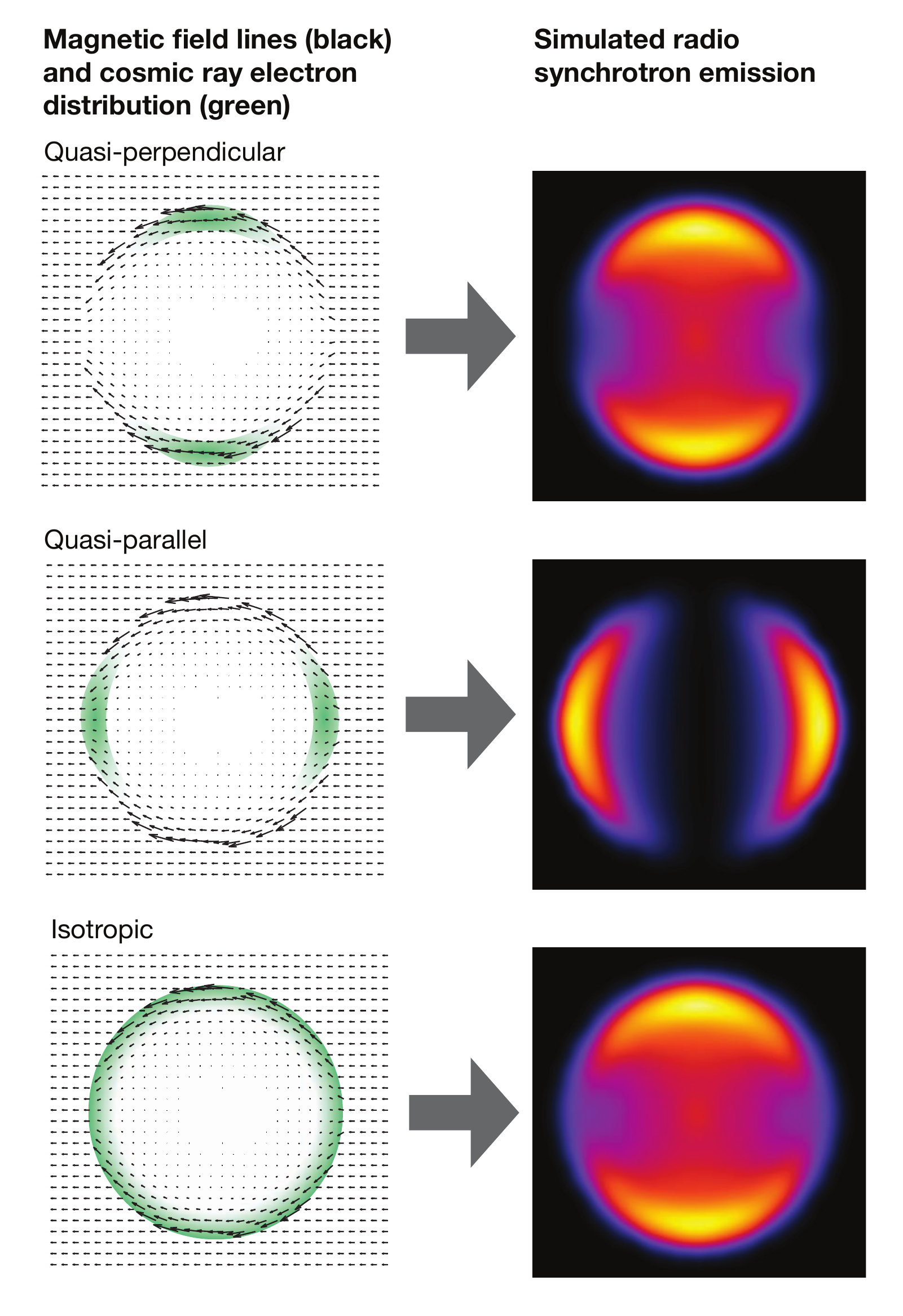}
\protect\caption{\label{fig:CRE-geometry} Geometry of CRE distributions for quasi-perpendicular shocks (top), quasi-parallel shocks (middle),
and the isotropic case (bottom) and the corresponding
simulated synchrotron emission, which has been normalized for display purposes. This cartoon is intended to qualitatively show the distribution of the CREs with respect to the magnetic field geometry. It is not intended to be representative of the precise quantitative distributions. }
\end{figure*}

\begin{figure*}
\centering
\includegraphics[width=12cm]{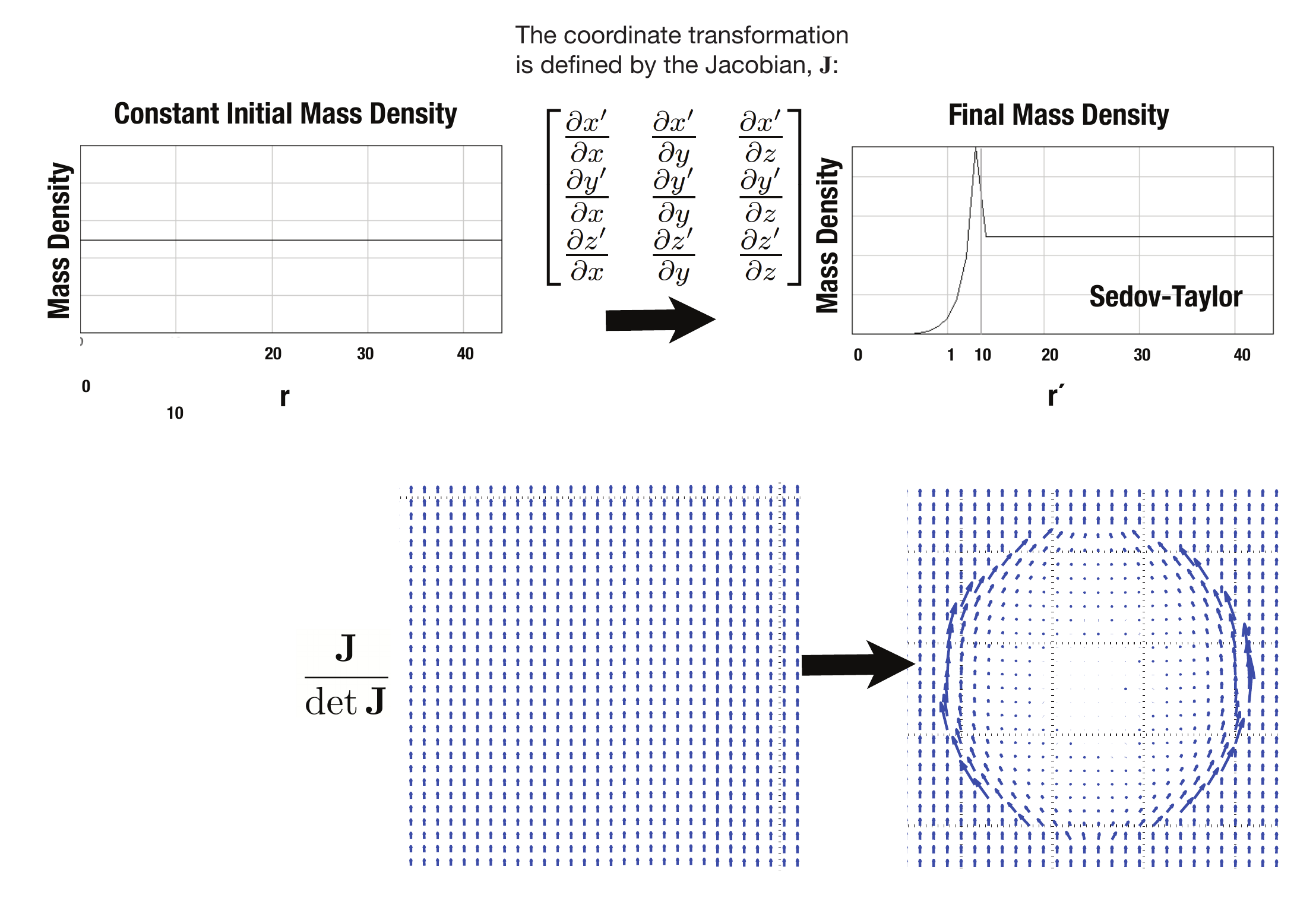}
\protect\caption{\label{fig:coord-transform}Illustration of the technique used to
determine the coordinate transformation matrix (Jacobian) and using
it to transform the magnetic field vectors. The model uses the values from the NE2001 thermal electron density model (both local to the SNR and elsewhere along the line of sight). The technique assumes that the initial thermal electron mass density is constant locally around the SNR, which is approximately true for the NE2001 model.} 
\end{figure*}

\begin{figure*}
\centering
\includegraphics[width=12cm]{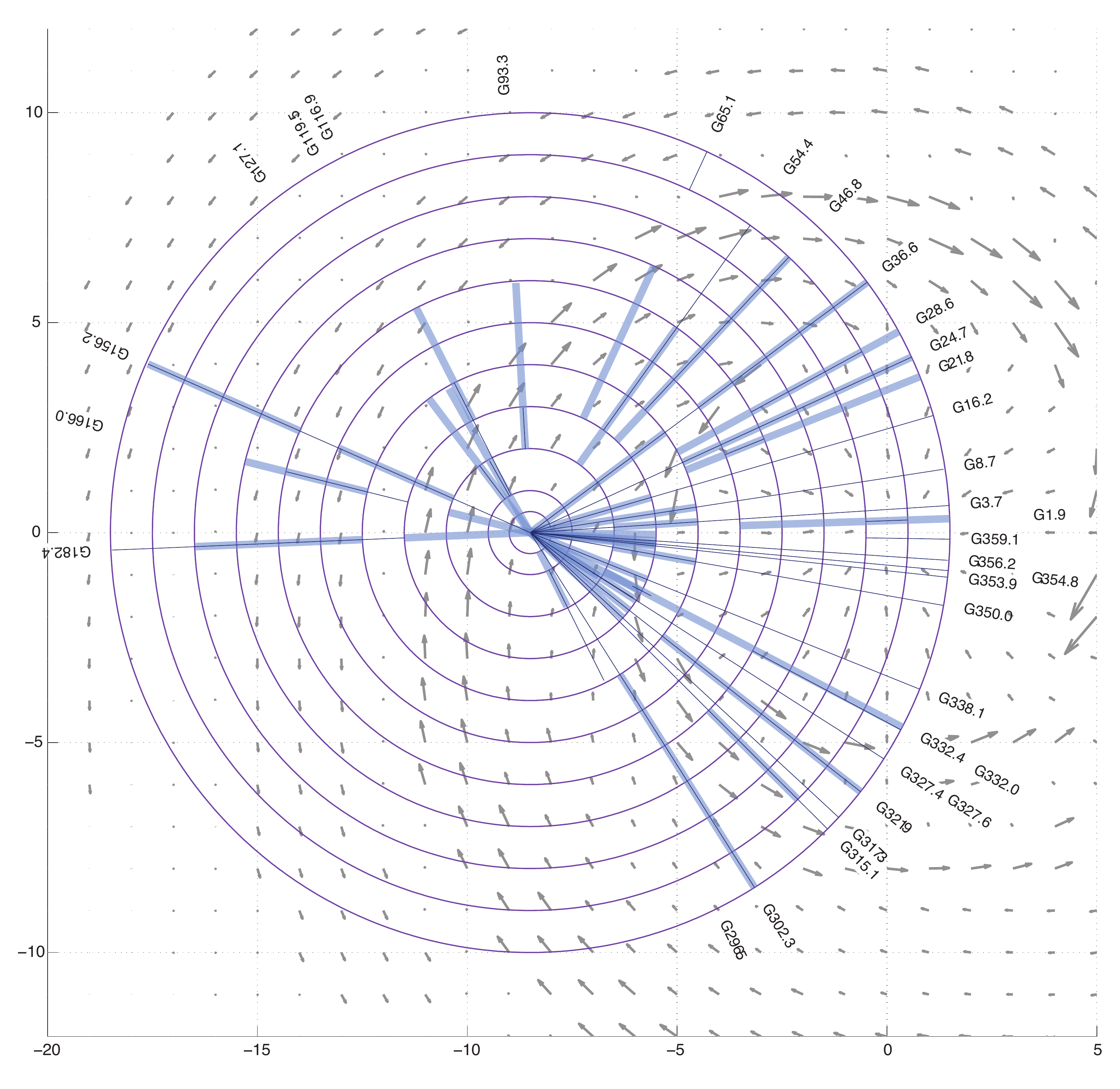}
\protect\caption{\label{fig:Galaxy-map}The position of the 33 SNRs in our sample showed
plotted on the Galaxy with field vectors from JF12
drawn for $z=0$, $b=0$. The Sun\textquoteright s position shown
is at \textminus 8.5~kpc from the Galactic centre. The thin lines
represent the distance estimate from the literature while the thicker,
lighter coloured lines represent the best fit distance from the models.}
\end{figure*}

\begin{figure*}
\centering
\includegraphics[width=17cm]{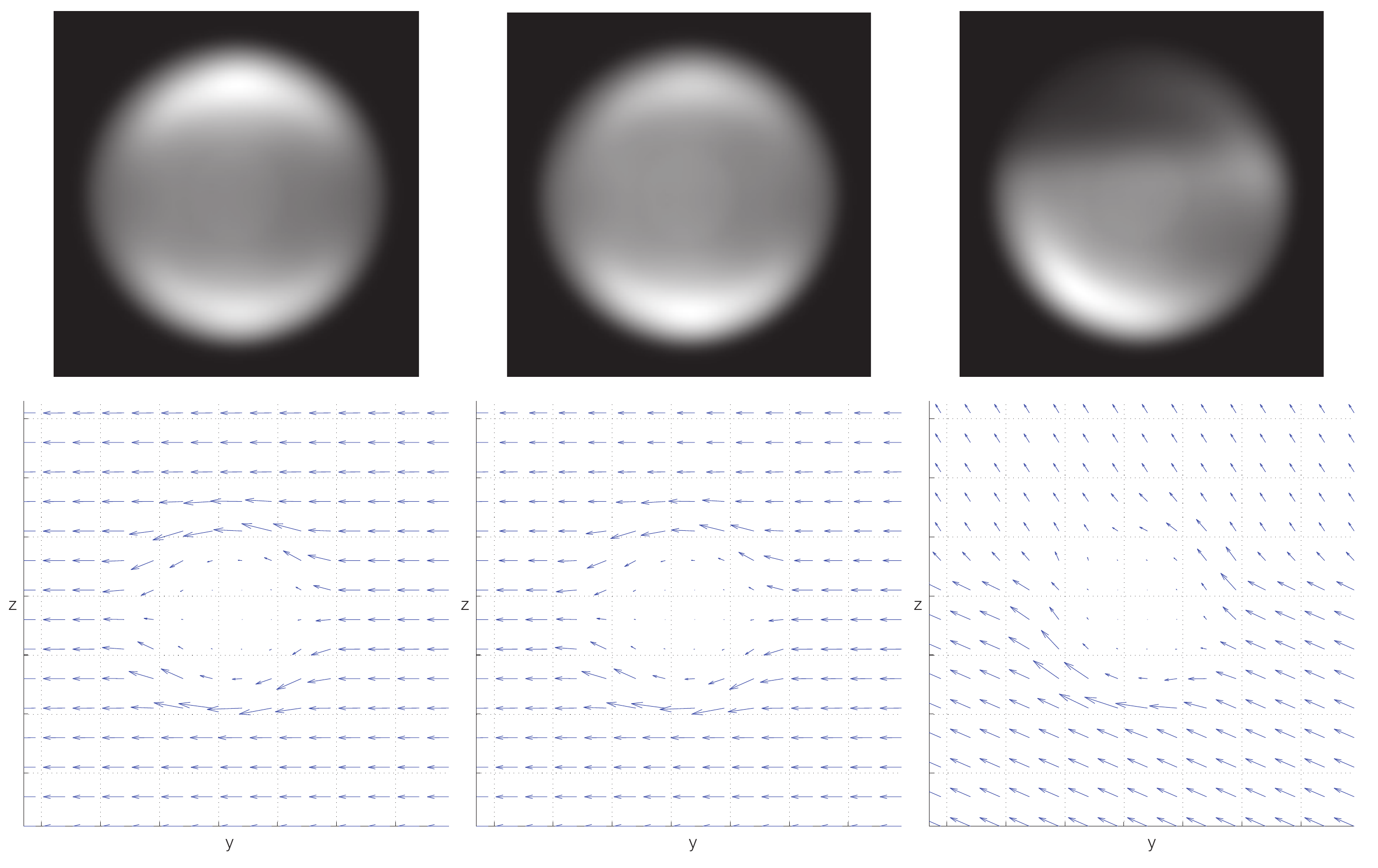}
\protect\caption{\label{fig:Top:-Models-at}{\bf Top}: Models at the position of G317.3--00.2
and for a distance of 2~kpc showing the impact of excluding the various
magnetic field components. {\bf Bottom}: The corresponding magnetic field
vectors shown in the $z$-$y$ plane (i.e. the plane of the sky, see Appendix
\ref{sub:Coordinate-System}) and cut through the centre of the SNR
bubble. {\bf Left panel}: Only the disk field, $B_{disk}$ has been included.
The two limbs are more or less uniform in brightness (slightly brighter
in the north). {\bf Centre panel}: The toroidal halo field, $B_{tor}$
has been included. There is now some asymmetry introduced since the
magnetic field is stronger in the south with the addition of the toroidal
halo component. {\bf Right panel}: The X-field, $B_{X-field}$, has now
been included. The vector sum of these components has now made the
field in the northern half of this SNR much smaller in magnitude and
the direction has changed (there is now a much stronger line-of-sight
($x$) component in the northern half, though that is not visible on
this figure). Thus the model SNR shows a high degree of asymmetry,
which strongly resembles the data.}
\end{figure*}

\begin{figure*}
\centering
\includegraphics[width=12cm]{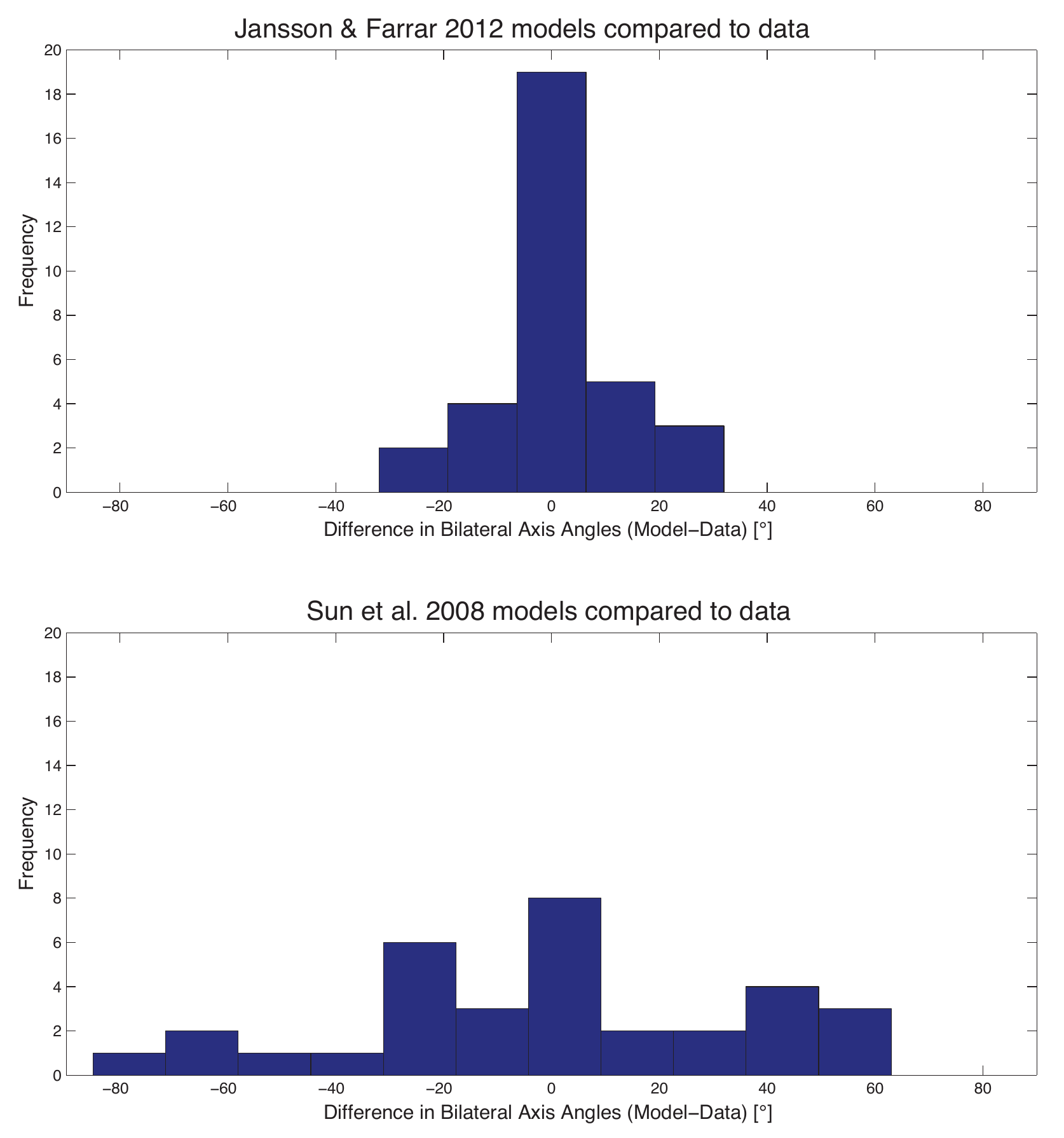}
\protect\caption{\label{fig:Histogram-of-the}Histogram of the differences, $\psi_{model}-\psi_{data}$, for the best fitting models of JF12 (top) and Sun et al. (2008) (bottom).
In the case of JF12, out of 33 SNRs, 25 have a difference that is $<10{^\circ}$, which
are in agreement within our uncertainty. In the case of Sun et al. (2008), only 10 SNRs have a difference that is $<10{^\circ}$.}
\end{figure*}

\begin{figure*}
\centering
\includegraphics[width=12cm]{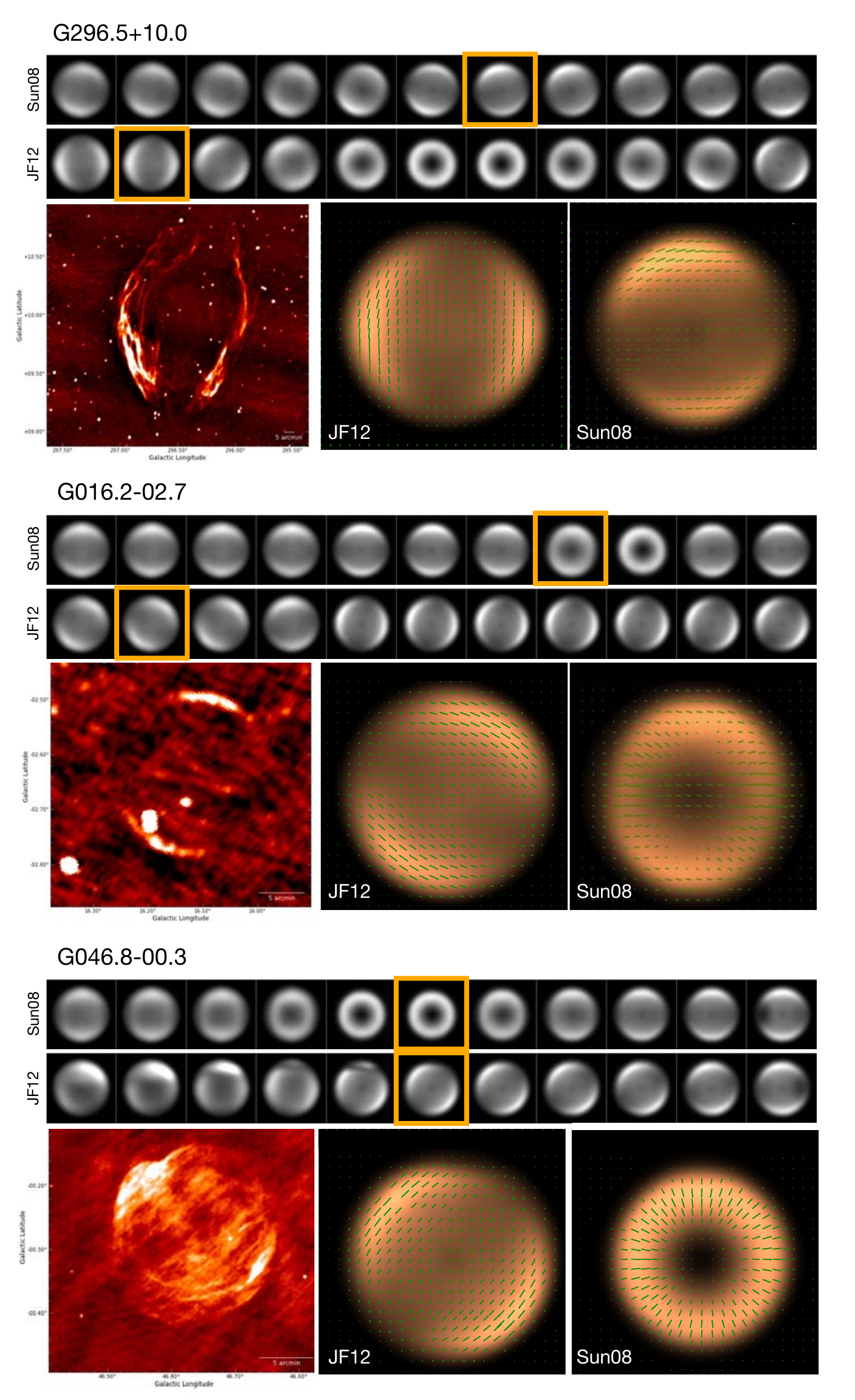}
\protect\caption{\label{fig:Comparison-of-SNR}Comparison of SNR models for three example SNRs: G296.5+10.0 (high-latitude), G016.2-2.7 (mid-latitude), and G046.8-0.3 (in the plane). 
All models from 0.5 to 10~kpc are shown (as described in Figure 5) for both GMF models: the Sun et al. (2008) (Sun08) and JF12. The orange box highlights the best fit in each case. Below the model strips we 
show the data (left) as well as the corresponding best fit model for JF12 (centre) and Sun08 (right) . The model polarization magnetic field vectors are overlaid in green. }

\end{figure*}
\begin{figure*}
\centering
\includegraphics[width=12cm]{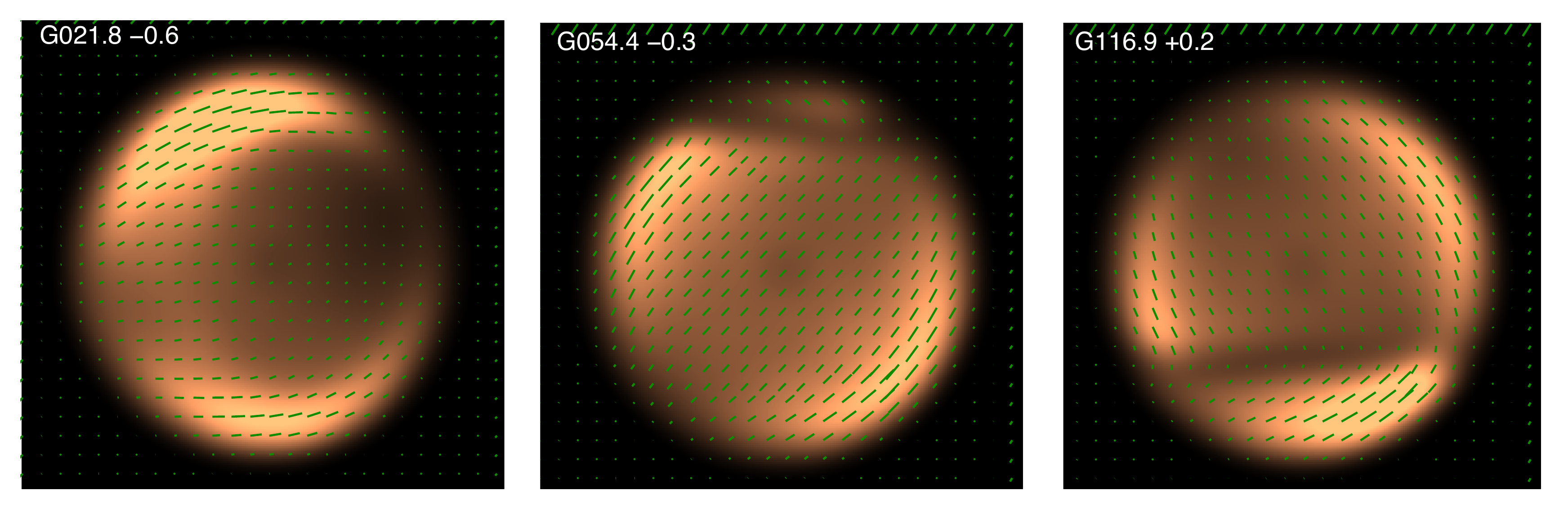}
\protect\caption{\label{fig:Simulated-polarization-mixed-B}Simulated polarization
vectors for SNRs where observations show that they have `mixed'
magnetic fields. These are all shown for a distance of 4~kpc, which
is the best-fit distance for G054.4--0.3 and G116.9+0.2. In the case
of G021.8--0.6, the best fit distance is at 5~kpc. The magnetic field at that
distance is tangential, but at 4~kpc, which is still a reasonable
fit, the field shows more characteristics of being `mixed'.}
\end{figure*}

\newpage

\onecolumn

\begin{table}

\caption{\label{tab:phys-ang-size}Summary of the physical and angular sizes
for the model SNRs bubbles at the various distances.}

\begin{tabular}{ccc}

\hline 
Distance (kpc) & Physical Radius (pc) & Angular Diameter ($'$)\tabularnewline
\hline 
\hline 
0.5 & 40 & 275\tabularnewline 
1 & 40 & 137\tabularnewline
2 & 40 & 69\tabularnewline
3 & 40 & 46\tabularnewline
4 & 60 & 52\tabularnewline
5 & 60 & 41\tabularnewline
6 & 60 & 34\tabularnewline
7 & 60 & 29\tabularnewline
8 & 60 & 26\tabularnewline
9 & 60 & 23\tabularnewline
10 & 60 & 21\tabularnewline
\hline 
\end{tabular}

\end{table}

\begin{table}
\caption{\label{tab:Number-of-SNRs}Number of SNRs\tablefootmark{a} with the various classifications
and of the various types.  }

\begin{tabular}{>{\centering}p{2cm}|>{\centering}p{1.8cm}>{\centering}p{1.8cm}>{\centering}p{1.9cm}|>{\centering}p{1.9cm}|c}
\hline 
\multirow{2}{2cm}{SNR type} & \multicolumn{3}{c|}{Axisymmetric\tablefootmark{b}} & \multirow{2}{1.9cm}{Not axisymmetric}  & \multirow{2}{*}{Total}\tabularnewline
 & Very clearly defined & Somewhat defined & Not clearly defined &  & \tabularnewline
\hline 
\hline 
Shell & 27 & 17 & 43 & 112 & 199\tabularnewline
\hline 
Plerionic-composite & 1 & 0 & 9 & 31 & 41\tabularnewline
\hline 
Thermal-composite & 4 & 3 & 2 & 27 & 36\tabularnewline
\hline 
Thermal \& plerionic-composite & 1 & 0 & 1 & 5 & 7\tabularnewline
\hline 
Unknown & 0 & 1 & 3 & 6 & 10\tabularnewline
\hline 
\hline 
Total & 33 & 21 & 58 & 181 & 293\tabularnewline
\hline 
\end{tabular}
\tablefoot{
\tablefoottext{a}{This table represents
the up-to-date numbers at the time of this writing, but the companion
website is dynamic and classifications and exact counts of SNRs of
various types are subject to change.}
\tablefoottext{b}{The `axisymmetric' SNRs include both
the so-called bilateral SNRs with two distinct limbs of emission and
also the one-sided shells. The morphology of the other SNRs are not consistent with this definition, thus we refer to these as `not-axisymmetric'. This includes SNRs with a 
round appearance, of which there are 43 candidates (33 shell, 7 plerionic composite, and 3 thermal composite). We consider only 15 of these candidates to be very clearly defined or somewhat defined, with the remainder (28 candidates) being not clearly defined. }
}
\end{table}

\begin{table}
\caption{\label{tab:data-vs-model}Summary of the results for the data with the corresponding
best fit model. For the `Ratio' column, N/S refers to the ratio
of the measurements of the North limb/South limb.}
\begin{tabular}{l|cc|cc|cc|c}
\multirow{2}{*}{SNR} & \multicolumn{2}{c|}{Distance [kpc]} & \multicolumn{2}{c|}{$\psi$ (\textpm{}5\textdegree{})} & \multicolumn{2}{c|}{Ratio: N/S} & \multirow{2}{1.0cm}{Image Ref.}\tabularnewline
 & Published & Model & Data & Model & Data & Model & \tabularnewline
\hline 
\hline 
G001.9+00.3 & 8.5 & $7_{-2}^{+4}$ & -85 & -89 & 1.0 & 1.0 & 1\tabularnewline
G003.7--00.2 & unknown & $3_{-1}^{+1}$ & 7 & 15 & 1.2 & 1.1 & 2\tabularnewline
G008.7--05.0 & unknown & $2_{-2}^{+1}$ & -35 & -35 & 1.0 & 1.1 & 3\tabularnewline
G016.2--02.7 & unknown & $0.5_{-0.5}^{+2.5}$ & -20 & -29 & 0.9 & 1.0 & 3\tabularnewline
G021.8--00.6 & 5.2-5.5 & $5_{-1}^{+6}$ & 36 & 61 & 0.6 & 1.1 & 4\tabularnewline
G024.7--00.6 & unknown & $7_{-3}^{+4}$ & -87 & 61 & 2.0 & 1.1 & 4\tabularnewline
G028.6--00.1 & 6-8.5 & $6_{-2}^{+5}$ & 63 & 79 & 0.8 & 0.9 & 4\tabularnewline
G036.6+02.6 & unknown & $2_{-2}^{+2}$ {[}or $8_{-3}^{+3}${]} & -24 & -35 & 1.4 & 1.7 & 3\tabularnewline
G046.8--00.3 & 4.3-8.6 & $5_{-2}^{+4}$ & 46 & 50 & 1.3 & 0.9 & 4\tabularnewline
G054.4--00.3 & 3.3-9 & $4_{-2}^{+2}$ & 47 & 50 & 2.0 & 0.9 & 5\tabularnewline
G065.1+00.6 & 9.0-9.6 & $5_{-2}^{+2}$ & 40 & 42 & 0.7 & 0.9 & 5\tabularnewline
G093.3+06.9 & 1.7-2.7 & $4_{-2}^{+2}$ & 6 & 11 & 0.8 & 0.6 & 6\tabularnewline
G116.9+00.2 & 1.6-3.5 & $4{}_{-1}^{+2}$ & -62 & -30 & 1.5 & 1.2 & 5\tabularnewline
G119.5+10.2 & 1.1-1.7 & $1_{-1}^{+3}$ & 36 & 11 & 0.5 & 0.9 & 7\tabularnewline
G127.1+00.5 & 1.1-1.3 & $1_{-1}^{+3}$ & 15 & 19 & 1.5 & 1.2 & 5\tabularnewline
G156.2+05.7 & 1.0-3.0 & $4_{-2}^{+1}$ {[}or $8_{-2}^{+3}${]} & -66 & -73 & 1.3 & 1.0 & 8\tabularnewline
G166.0+04.3 & 3-6 & $5_{-1}^{+2}$ {[}or $1_{-1}^{+1}${]} & 0 & 3 & 1.8 & 0.8 & 5\tabularnewline
G182.4+04.3 & >3 & $6_{-2}^{+2}$ {[}or $1_{-1}^{+2}${]} & -17 & 3 & 0.5 & 0.7 & 5\tabularnewline
G296.5+10.0 & 1.3-3.9 & $1_{-0.5}^{+1}$ & 81 & 80 & 1.1 & 1.1 & 9\tabularnewline
G302.3+00.7 & unknown & $7_{-3}^{+3}$ & 45 & 50 & 1.2 & 1.0 & 9\tabularnewline
G315.1+02.7 & unknown & $7_{-1}^{+2}$ & 65 & 67 & 1.1 & 1.0 & 10\tabularnewline
G317.3--00.2 & unknown & $1_{-1}^{+2}$ & -35 & -40 & 0.6 & 0.3 & 9\tabularnewline
G321.9--00.3 & unknown & $8_{-4}^{+3}$ {[}or $1_{-1}^{+2}${]} & -35 & -50 & 1.3 & 1.1 & 9\tabularnewline
G327.4+01.0 & unknown & $1_{-0.5}^{+2}$  & -40 & -44 & 3.7 & 0.9 & 9\tabularnewline
G327.6+14.6 & 1.6-2.2 & $1_{-0.5}^{+1}$ & 83 & 79 & 1.0 & 1.1 & 9\tabularnewline
G332.0+00.2 & >6.6 & $1{}_{-1}^{+2}$ & -10 & -11 & 0.9 & 0.3 & 9\tabularnewline
G332.4--00.4 & 3.4 & $3_{-1}^{+8}$  & 11 & 15 & 0.9 & 1.1 & 9\tabularnewline
G338.1+00.4 & unknown & $2_{-2}^{+1}$  & -57 & -40 & 0.4 & 0.8 & 9\tabularnewline
G350.0--02.0 & unknown & $3_{-1}^{+1}$  & 6 & 15 & 4.1 & 1.2 & 10\tabularnewline
G353.9--02.0 & unknown & $1_{-1}^{+2}$  & -33 & -25 & 0.8 & 1.0 & 3\tabularnewline
G354.8--00.8 & unknown & $1_{-1}^{+2}$  & -21 & -25 & 1.0 & 0.9 & 9\tabularnewline
G356.2+004.5 & unknown & $1_{-1}^{+2}$  & -34 & -40 & 1.0 & 0.9 & 3\tabularnewline
G359.1--00.5 & 8-10.5 & $1_{-1}^{+2}$  & -36 & -34 & 0.8 & 0.7 & 9\tabularnewline
\hline 
\end{tabular}

\tablebib{(1) \citet{Reynolds:2008fy}, (2) Very Large Array via NRAO Science
Data Archive, (3) The NRAO VLA Sky Survey (NVSS, \citealp{Condon:1998kn}),
(4) MAGPIS: A Multi-Array Galactic Plane Imaging Survey \citep{Helfand:2006kw},
(5) Canadian Galactic Plane Survey, (CGPS, \citealp{Taylor:2003iz}),
(6) \citet{Landecker:1999im}, (7) The Westerbork Northern Sky Survey
(WENSS, \citealp{Rengelink:1997fo}), (8) Sino-German $\lambda6$
cm polarization survey \citep{Gao:2010fu}, (9) The Molonglo Observatory
Synthesis Telescope (MOST) Supernova Remnant Catalogue, \citep{Whiteoak:1996tn},
(10) The Parkes-MIT-NRAO surveys \citep{Condon:1993kf}.\\
Published distances are taken from SNRcat \citep{Ferrand:2012cr} and references therein.
}
\end{table}

\end{document}